\documentclass[a4paper,12pt]{article}
\usepackage{amsfonts}
\usepackage{mathrsfs}
\usepackage{amsmath}
\usepackage{amssymb}
\usepackage[medium]{titlesec}
\usepackage{bm}
\usepackage{cite}
\usepackage[normalem]{ulem}
\usepackage{extarrows}
\usepackage{slashed}
\usepackage{isodateo}
\usepackage{graphicx}
\usepackage{xcolor}
\usepackage[bookmarksnumbered=true,bookmarksopen=true]{hyperref}
 \hypersetup{colorlinks,%
             linkcolor=[rgb]{0,0.2,0.6}, %
             citecolor=[rgb]{0,0.2,0.6}}
\usepackage[hmargin=.7in,vmargin=1.1in]{geometry}
\usepackage{indentfirst}
\newcommand{\FR}[2]{\displaystyle\frac{\,{#1}\,}{#2}}

\newcommand{\n}{\nonumber}

\graphicspath{{fig/}}

\def\bge{\begin{equation}}
\def\ede{\end{equation}}
\def\bga{\begin{aligned}}
\def\eda{\end{aligned}}
\def\bgp{\begin{pmatrix}}
\def\edp{\end{pmatrix}}
\def\bgs{\begin{subequations}}
\def\eds{\end{subequations}}
\newcommand{\order}[1]{\mathcal{O}({#1})}
\def\di{{\mathrm{d}}}

\def\mb{\mathbf}

\def\pd{\partial}

\def\la{\langle}\def\ra{\rangle}

\def\to{\rightarrow}
\def\To{\Rightarrow}

\def\al{\alpha}
\def\be{\beta}
\def\ga{\gamma}

\def\ep{\epsilon}

\def\rh{\rho}
\def\si{\sigma}

\newcommand{\ob}[1]{\mkern 2mu \overline{\mkern -2mu #1 \mkern -2mu}\mkern 2mu}

\newcommand{\wh}[1]{\mkern 2mu \widehat{\mkern-2mu#1\mkern-2mu}\mkern 2mu}

\begin{document}
%\begin{fmffile}{ }
%\fancyhf{}
%\lhead{}
%\rhead{Begin on 2017/04/16, last updated on \TeXdate\today}
%\cfoot{\thepage}

\title{ \Large\textbf{An Analytical Portrait of Binary Mergers in\\ Hierarchical Triple Systems}}
\author{Lisa Randall$^a$\footnote{Email: randall@physics.harvard.edu} ~and~ Zhong-Zhi Xianyu$^{a,b}$\footnote{Email: xianyu@cmsa.fas.harvard.edu}\\[2mm]
\normalsize{$^a$~\emph{Department of Physics, Harvard University, 17 Oxford St., Cambridge, MA 02138, USA}}\\
\normalsize{$^{b}$~\emph{Center of Mathematical Sciences and Applications, Harvard University,}} \\
\normalsize{\emph{20 Garden St., Cambridge, MA 02138, USA}}}
\date{}
\maketitle

\vspace{3cm}

\begin{abstract}

With better statistics and precision, eccentricity could prove to be a useful tool for understanding the origin and environment of binary black holes. Hierarchical triples in particular, which might be abundant in globular clusters and galactic nuclei, could generate observably large eccentricity at LIGO and future gravitational wave detectors. Measuring the eccentricity distribution accurately could help us probe the background and the formation of the mergers. In this paper we continue our previous investigation and improve our semi-analytical description of the eccentricity distribution of mergers hierarchical triple systems. Our result, which further reduces the reliance on numerical simulations, could be useful for statistically distinguishing different formation channels of observed binary mergers.

\end{abstract}

\newpage

\section{Introduction}
\label{sec_intro}

Since the first LIGO detection of gravitational waves (GWs) from the merger of a pair of black holes (BHs) \cite{Abbott:2016blz}, five additional detections have been announced \cite{Abbott:2016nmj,Abbott:2017vtc,Abbott:2017gyy,Abbott:2017oio,TheLIGOScientific:2017qsa}. The expectation is that many more will come with the promise of increasing duration and observational sensitivity. We can begin to ask what we will learn from the mergers aside from the properties of the black holes and neutron stars themselves. In particular we ask whether features of the detections can shed light on the formation channels and even teach us the ambient matter distribution of the mergers. 
 
The literature suggests several ways that such mergers might form and also provides estimates of their rates.  See \cite{Abadie:2010cf} for a review. In a galaxy, binary BHs (BBHs) can form in less dense regions, in which case they are isolated binaries. Isolated binaries receive little perturbation from the ambient matter. For such BBHs, the hardening mechanism before entering the LIGO sensitivity window is due almost exclusively to GW emission. However, it is well known that GW is very efficient in circularizing the orbits through energy and angular momentum reduction. The orbits of these BBHs will have measurable eccentricity if and when they enter the LIGO  band only if formed with a significant natal kick.

On the other hand, BBHs can also form in  denser regions in a galaxy such as in the nuclear cluster (NC) at the center of a galaxy and in globular clusters (GCs), in what is known as dynamical formation scenarios \cite{Bartos:2016dgn,McKernan:2017umu,Leigh:2017wff}. Higher ambient density can affect BBH formation in a number of ways, potentially generating observably large eccentricity for the orbits of BBHs, in contrast to isolated BBHs. The eccentricity in this case could be generated by persistent perturbation from a third body, or by occasional direct closed two-body encounters. 

Observational rates for the different merger channels have been estimated theoretically, though with rather large uncertainties (typically two or even three orders of magnitude).   In field binaries, the poorly constrained natal kick might be the source of ellipticity whereas in dynamical formation scenarios, the natal kicks could eject BHs out of star clusters if they are higher than the escape velocity. Therefore stronger natal kicks could suppress the formation of BBHs in GCs, which  typically have  smaller escape velocity than a NC \cite{Antonini:2016gqe} so their influence could in principle either enhance or suppress measured eccentricities.
The binary population sufficiently close to a central black hole in galactic nuclei is also unclear but formation scenarios in a disk can potentially lead to interesting black hole binary populations near the central region \cite{Volonteri:2007dx}.

Such estimates do suggest that dynamically formed binaries in dense environments could contribute significantly to observable events. It is worth investigating whether a better understanding of parameter distributions and their statistical correlation could help us disentangle different formation channels. A number of studies have focused on this problem, studying the mergers in GCs, NCs with or without a central supermassive BH (SMBH), most of which have relied on Monte-Carlo simulations \cite{Gultekin:2005fd,Miller:2008yw,Antonini:2012ad,Stephan:2016kwj,VanLandingham:2016ccd,Antonini:2016gqe,Silsbee:2016djf,Antonini:2017ash,Petrovich:2017otm,Hoang:2017fvh,Gondan:2017wzd,Giersz:2015,Bartos:2016dgn,Askar:2016jwt}. 

In any case an analytical understanding of the evolution of binary mergers, and in particular, of its eccentricity generation, will help us  more efficiently connect the initial distribution of binaries with the final distribution of eccentricity in the observational window. This approach can also provide us a better physical intuition for the parameter dependence of the mergers, and thus help us to better understand the qualitative difference of different formation channels.

The Kozai-Lidov solutions with GW back-reaction and post-Newtonian (PN) corrections to the inner binary have been extensively studied in the literature \cite{Miller:2002pg,Blaes:2002cs,Wen:2002km,Naoz:2011mb}. In this paper we will describe how to estimate analytically the final eccentricity of binary mergers at the LIGO threshold for such solutions with arbitrary initial conditions for the relevant parameters. This will provide a clearer understanding of the parameter dependence and make it faster to study the statistical correlation between various parameter distributions, as compared with Monte Carlo simulations. 

In \cite{Randall:2017jop} we initiated a semi-analytic study of the induced eccentricity for dynamically formed binaries, with a focus on binaries that we assumed were formed in the vicinity of a SMBH \cite{Bahcall:1976aa}. This analysis will assume that there are binary black holes close enough to the central massive black hole for the KL mechanism to be effective. Although the distribution is not known and it is possible black holes are removed through dynamical effects, situations with a disk for example \cite{Morris:1993zz,Bartos:2016dgn} can cause the black holes to migrate to the central region of an AGN through dynamical friction. The binary with this SMBH formed a quasi-stable triple system, known as a hierarchical triple. The binary mergers in hierarchical triples  can be elliptical enough to be observed  even as they enter the LIGO sensitivity band, despite the circularization via GWs. In a triple system, the eccentricity can  be generated either from secular exchange of angular momentum between the ``inner'' binary and the larger system, known as the KL mechanism \cite{Kozai:1962zz,Lidov1976}, or be quickly generated in non-perturbative solutions \cite{Randall:2017jop}. These two mechanisms generate very different eccentricity distributions and also apply in different situations. For hierarchical triples in galactic centers, higher order multipoles are suppressed by the ratio of orbital sizes which is smaller than a percent. This is further justified by $N$-body simulation, which have been shown to agree with the leading order expansion in \cite{Antonini:2012ad} when considering orbits near galactic centers. This approximation breaks down for triples in globular clusters, where the perturbing object is much closer and octupole and higher order terms can play a big role \cite{Antonini:2015zsa}. We leave further analytical work with higher order terms to further study. 

In \cite{Randall:2017jop} the two competing effects of KL oscillation and GW circularization were explicitly taken into account. There we showed that the evolution of binary mergers in hierarchical triples comes with two qualitatively distinct stages depending on the relative change in eccentricity due to the KL effect and to GW emission. In the first stage, the KL oscillation is strong enough to generate large eccentricity while the GW radiation slowly reduces the binary separation without significantly erasing the eccentricity significantly. For smaller binary separations, the KL oscillations get weaker while the GW radiation grows stronger. The binary separation eventually becomes so small that the change in eccentricity due to GW emission dominates over the change due to KL oscillations. This is the starting point of the second stage, during which the eccentricity decreases monotonically with time until entering  the observational band.

We showed in \cite{Randall:2017jop} how to analytically calculate the final eccentricity in the LIGO band by accounting for both the KL effect and GW emission up to a background-dependent cut-off on the distance to the central SMBH, and a distribution of eccentricity $f(e)$ characterizing the beginning of the second stage. The cut-off is a consequence of requiring the binaries to merge before they are evaporated through scattering with background stars \cite{Leigh:2017wff}. Both of these two unknowns can be calculated only by explicit inclusion of the post-Newtonian (PN) corrections to the binary evolution.  The leading PN effect, i.e., the precession of the binary orbit, tends to destroy the KL oscillation and thus reduces the maximal eccentricity the binary can reach in each KL period. This maximal eccentricity of each cycle affects both $f(e)$ at the onset of the second stage and the merger time of the binary, and thus the cut-off. 

In this paper we improve on our previous study by systematically including  all three effects: KL, GW, and PN -- the latter of which we absorbed in numerically determined parameters in \cite{Antonini:2012ad}. We will calculate analytically the cutoff that was previously determined numerically, and our method here will bypass the need for determining $f(e)$ at the boundary between the two stages. We present the analysis for a general hierarchical triple system so that it can be applied not only to binaries near a SMBH but also to other environments such as GCs.

Using our analytical estimate, we calculate the eccentricity distribution of BBH mergers in SMBH-carrying NCs, and how it depends on the density profile of BBHs. For example, we show that the final eccentricity in this channel is anti-correlated with the binary mass, which could be useful for disentangling this formation channel from others in which the opposite correlation applies. We also show that the eccentricity depends on the background density profile in NCs. This dependence is potentially important for a better understanding of the mass distribution in the vicinity of a SMBH. While it is known theoretically that a fully relaxed NC with central SMBH would develop a Bahcall-Wolf cusp \cite{Bahcall:1976aa}, no observational evidence is known for such a cusp profile. In fact, in less relaxed galaxies, the density profile around the SMBH could be flatter. Furthermore, DM could also play a role in the formation of the background. Since it is extremely difficult to resolve the central region of a NC, well-measured BBH mergers might turn out to be a unique probe into this densest environment in a galaxy.

In Sec.~\ref{sec_triple} we review the hierarchical triple system and the secular evolution of its orbital parameters, taking account the GW back-reaction and the PN correction to the small orbit. We present our analytic method to estimate the merger time and final eccentricity of the inner binary for arbitrary initial parameters in Sec.~\ref{sec_merger}, along with several typical numerical solutions to the equations of secular evolution. In Sec.~\ref{sec_ED} we characterize the eccentricity distribution of mergers in SMBH-carrying NCs. Further discussions are in \ref{sec_discussions}. In Appendix~\ref{app_small_e} we estimate the LIGO sensitivity to a small eccentricity.

\section{Review of Hierarchical Triple Systems}
\label{sec_triple}

In this section we derive a set of equations governing the long-term evolution of a binary system, taking into account the three dominant effects: 1) tidal perturbation from a tertiary body; 2) PN precession of the orbit; 3) GW back reaction. The equations for these effects are well known and widely applied in the literature. A nice presentation of the KL mechanism can be found in \cite{Lidov1976}. An introduction to PN precession can be found in any standard textbook of general relativity, e.g., \cite{Weinberg:1972kfs}. Finally, the back reaction of GW emission on an elliptic binary orbit was firstly studied in \cite{Peters:1964zz}. Here we review the equations describing these three processes along with hierarchical triple systems for readers not familiar with the field. We first review the geometric configuration of a hierarchical triple system, introducing the notation and terminology. We then derive the Hamiltonian and equations of orbital evolution. 

\subsection{The Hierarchical Triple}

We assume three bodies with masses $m_0$, $m_1$, and $m_2$. In an inertial frame, the positions of the three bodies are described by three vectors, $\mb r_0$, $\mb r_1$, and $\mb r_2$, respectively. With Newtonian gravitation, the Hamiltonian of the system is given by,
\bge
\label{Ham}
  H=\FR{1}{2m_0}|\mb p_0|^2+\FR{1}{2m_1}|\mb p_1|^2+\FR{1}{2m_2}|\mb p_2|^2-\FR{Gm_0m_1}{|\mb r_0-\mb r_1|}-\FR{Gm_0m_2}{|\mb r_0-\mb r_2|}-\FR{Gm_1m_2}{|\mb r_1-\mb r_2|},
\ede
where $\mb p_i=m_i\dot{\mb r}_i~(i=0,1,2)$ is the momentum conjugate to $\mb r_i$, and $G$ is the gravitational constant, which equals to $4\pi^2$ in astronomical units (AU, yr, $M_\odot$).

It is well known that the Keplerian two-body problem can be recast into two independent motions, namely the (trivial) motion of the mass center and the motion of the reduced mass relative to the mass center. The three-body problem considered here can be treated similarly. We group $m_0$ and $m_1$ as a binary system, called the ``inner'' binary. The inner binary and $m_2$ then form an effective two-body system, called the ``outer'' binary. Thus, we introduce the following set of variables,
\begin{align}
\label{mass}
  &m\equiv m_0+m_1,
  &&M\equiv m_0+m_1+m_2,
  &&\mu_1\equiv \FR{m_0m_1}{m_0+m_1},
  &&\mu_2\equiv \FR{(m_0+m_1)m_2}{m_0+m_1+m_2},
\end{align}
which correspond to the total masses of the inner binary, of the outer binary, and the reduced masses of the inner binary, and of the outer binary, respectively. We also introduce the following position vectors,
\begin{align}
  &\mb R=\FR{1}{M}(m_0\mb r_0+m_1\mb r_1+m_2\mb r_2),
  &&\mb R_1=\mb r_1-\mb r_0,
  &&\mb R_2=\mb r_2-\FR{1}{m}(m_0\mb r_0+m_1\mb r_1),
\end{align}
and the corresponding momenta,
\begin{align}
  &\mb P=M\dot{\mb R},
  &&\bm \Pi_1=\mu_1\dot{\mb R}_1,
  &&\bm \Pi_2=\mu_2\dot{\mb R}_2.
\end{align}
Then it is easy to see that the Hamiltonian (\ref{Ham}) can be rewritten as,
\bge
\label{Hsep}
  H=\FR{1}{2M}|\mb P|^2+\bigg(\FR{1}{2\mu_1}|\mb \Pi_1|^2-\FR{Gm\mu_1}{|\mb R_1|}\bigg)+\bigg(\FR{1}{2\mu_2}|\mb \Pi_2|^2-\FR{GM\mu_2}{|\mb R_2|}\bigg)+H',
\ede
where the system breaks down two three ``independent'' motions, namely the motion of the mass center of the triple (which is trivial and will be neglected), and of the inner and outer binary, plus the ``interaction'' between the inner and outer binary described by $H'$,
\bge
  H'=\FR{Gmm_2}{|\mb R_2|}-\FR{Gm_0m_2}{|\mb R_2+\frac{m_1}{m}\mb R_1|}-\FR{Gm_1m_2}{|\mb R_2-\frac{m_0}{m}\mb R_1|}.
\ede
Assuming that the inner and outer binaries are ``weakly coupled'', we can perform a perturbative multipole expansion of $H'$ at position $\mb R_2$ as follows,
\begin{align}
  &H'=\sum_{\ell=2}^{\infty} H^{(\ell)}, &&H^{(\ell)}=-\FR{Gm_0m_1m_2}{m^\ell}\big[m_0^{\ell-1}+(-1)^\ell m_1^{\ell-1}\big]\FR{R_1^\ell}{R_2^{\ell+1}}\text{P}_\ell(\cos\varphi),
\end{align}
where $\varphi$ is the angle between $\mb R_1$ and $\mb R_2$, and $\text{P}_\ell(z)$ is Legendre polynomial. The leading nonvanishing term is the quadrupole interaction with $\ell = 2$, 
\bge
\label{Hquad}
  H^{(2)}=-\FR{Gm_0m_1m_2}{2m}\FR{R_1^2}{R_2^3}(3\cos^2\varphi-1),
\ede
which will be our main focus. To quantify the meaning of the weak coupling, we compare the magnitude of $H^{(2)}$ with the Hamiltonian of the  two binaries, i.e., the terms in the two big parentheses in (\ref{Hsep}). It is easy to see that the conditions for weak coupling between the inner and outer binaries are,
\begin{align}
\label{weakcondition}
  & \FR{m_2}{m}\bigg(\FR{R_1}{R_2}\bigg)^3 \ll 1, && \FR{\mu_1}{m}\bigg(\FR{R_1}{R_2}\bigg)^2\ll 1.
\end{align}
Therefore, in the weak coupling regime, we can retain the  quadrupole term (\ref{Hquad}) only, and study its perturbation on the motion of the two binaries. Before doing so, in the next subsection we will introduce orbital parameters that are more suitable for perturbation theory.

\subsection{Orbital Parameters and Delauney Variables}

In the Hamiltonian (\ref{Hsep}) the motion of the triple breaks down to two Keplerian two-body orbits, plus the perturbative interaction between them. It is thus useful to review the standard orbital parameters describing the configuration and orientation of the two-body orbit. In this subsection we focus on one orbit only and thus drop the subscript $(1,2)$ distinguishing the inner and outer orbits.

\begin{figure}[tbph]
\centering
\includegraphics[width=0.55\textwidth]{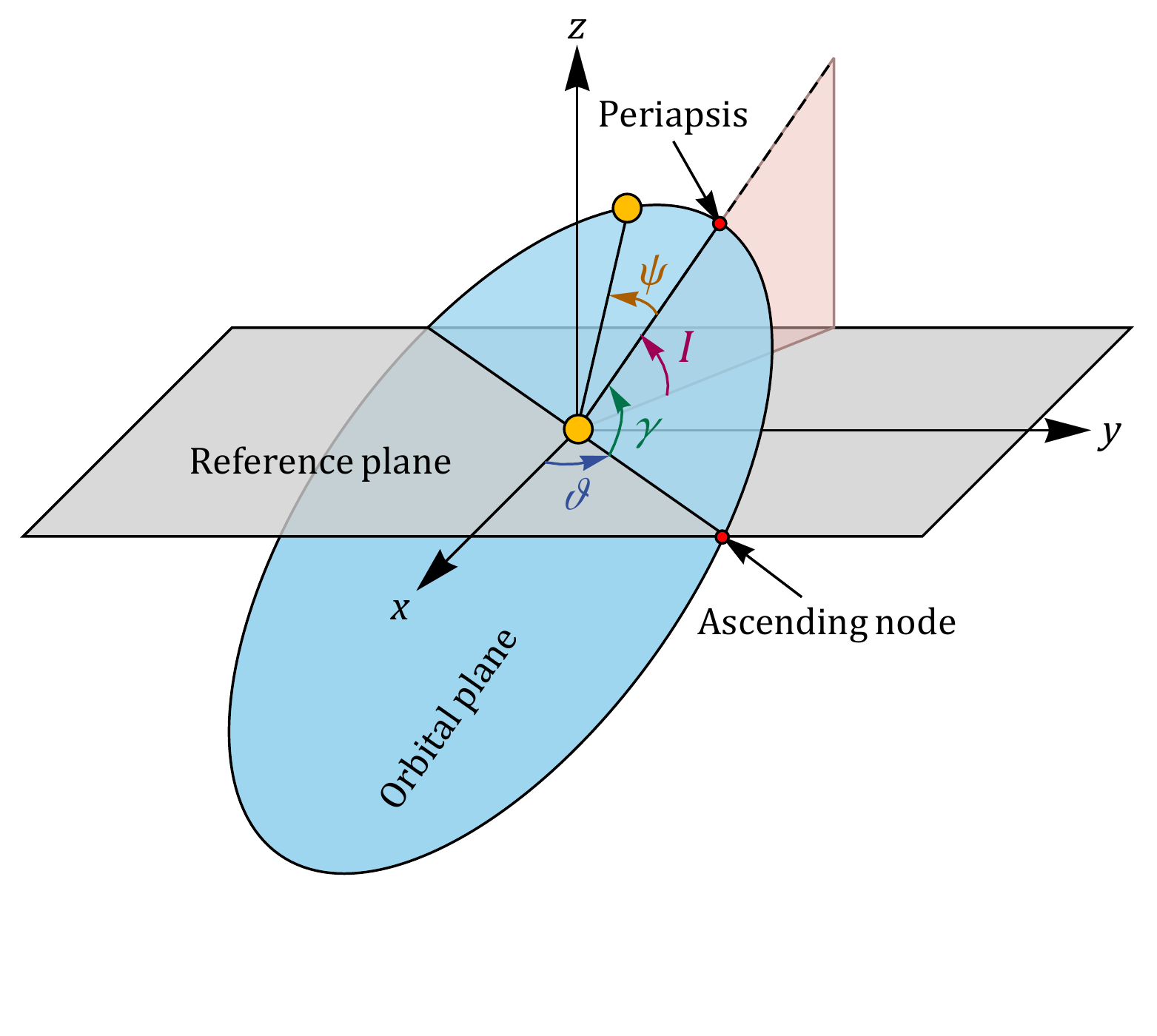}\\[-13mm]
\caption{An illustration of orbital parameters.}
\label{Fig_orbit}
\end{figure}

Since we are concerned only with bound orbits, which are always elliptical, the size and the shape of the orbit can always be described by two parameters, namely the \emph{semi-major axis} $a$ and the \emph{eccentricity} $e$. For example, the semi-minor axis $b$ is given by $b=a\sqrt{1-e^2}$, and the distance from the periapsis to the focus is $a(1-e)$. Then, the location of the rotating body in the orbital plane can be determined by one additional parameter $\psi$ called the \emph{true anomaly}, which is defined to be the angle from the periapsis to the rotating body in the orbital plane, as shown in Fig.\;\ref{Fig_orbit}. In addition to $(a,e,\psi)$, we need three Euler angles to characterize the orientation of the orbit, relative to some reference plane, also shown in Fig.\;\ref{Fig_orbit}. Conventionally, the three angles are chosen to be $(\vartheta,I,\ga)$, where $\vartheta$ is the angle from the reference direction to the ascending node on the reference plane, and is called the \emph{longitude of ascending node}, $I$ is the angle between the reference plane and the orbital plane, called the \emph{inclination}, and finally, $\ga$ is the angle from the ascending node to the periapsis in the orbital plane and is called \emph{the argument of periapsis}. The six-parameter set $(a,e,\psi,\theta,I,\ga)$ then completely characterize the position of the rotating body. 

In addition to these, we also use alternative widely-used parameters in the following. First, the \emph{mean anomaly} $\be$ and the \emph{eccentric anomaly} $u$ are related to the true anomaly $\psi$ through the following relations,
\begin{align}
\label{anomaly}
  &\be=u-e\sin u, && \cos\psi=\FR{\cos u-e}{1-e\cos u}.
\end{align}
The mean anomaly $\be$ undergoes periodic motion and has the same period as the true anomaly $\psi$, but it is defined such that the motion is uniform in time, namely $\be=\omega t$, with the orbital frequency $\omega^2=Gm/a^3$. The eccentric anomaly $u$ makes the relation between the true anomaly $\psi$ and the mean anomaly $\be$ explicit. Furthermore, since $1-e^2$ is a frequently appearing combination, it is useful to define it as a new parameter $\ep$,
\bge
  \ep\equiv 1-e^2.
\ede

The Hamiltonian (\ref{Hsep}) for the triple system is expressed in terms of the coordinates of the three bodies and their conjugate momenta. While this set of coordinates is intuitively simple, they are less suitable for our study here since eventually we want to trace the slow variation of orbital parameters rather than the exact locations of the three bodies. For systems involving librations or periodic motions like ours, there are well-known angle-action variables that are in place for our purpose. In celestial mechanics, a widely used set of angle-action variables are Delauney variables. For a Keplerian two-body system, the three angle variables $(\be,\ga,\vartheta)$ are given by the mean anomaly $\be$, the argument of the periapsis $\ga$, and the longitude of ascending node $\vartheta$, respectively. The three conjugate action variables $(J_\be,J_\ga,J_\vartheta)$ are given by,
\begin{align}
\label{actionvariable}
  &J_\be=\mu\sqrt{Gma}, &&J_\ga=\mu\sqrt{Gma(1-e^2)}, &&J_\vartheta=\mu\sqrt{Gma(1-e^2)}\cos I,
\end{align}
where $m$ is the total mass of the binary, $a$ is the semi-major axis of the orbit, $e$ is the eccentricity, and $I$ is the inclination. We note that the grouping of angle and action variables mixes up the position and orientation variables. The three angle variables include one position variable (the mean anomaly) and two Euler angles ($\ga,\theta$), while the third Euler angle $I$ is in the action variable $J_\vartheta$. We also note that $J_\ga$ coincides with the magnitude of the angular momentum, while $J_\vartheta$ is the component of the angular momentum orthogonal to the reference plane.
In terms of the Delauney variables, the Hamiltonian of an isolated binary becomes,
\bge
\label{Hbi}
  H_\text{binary}=-\bigg(\FR{Gm}{J_\be}\bigg)^2.
\ede 
The advantage of the Delauney variables can now be understood by noting that the Hamiltonian has explicit dependence on $J_\be$ only, which means that all Delauney variables but $\be$ are conserved for an isolated binary system, while $\be$ is under periodic motion for an elliptic orbit. After introducing a perturbation, the rest of the variables should undergo only slow motion, and we shall derive the corresponding equations in the next subsection.

\subsection{The equations of secular evolution}

The equations governing the long-term evolution of various orbital elements have three types of perturbations, namely the tidal perturbation at  quadrupole order, the PN precession of the orbit, and the back-reaction of the GW radiation. We consider these three pieces in turn.

\subsubsection{Tidal perturbation at quadrupole order}

As mentioned earlier, an isolated Keplerian binary system has an elliptical orbit with fixed orbital parameters $(a,e)$ and orientation. After turning on a perturbation, those parameters could undergo slow time variation. Specifically, for the hierarchical triple described by the Hamiltonian (\ref{Hsep}), we can rewrite the two binary terms in big parentheses in Delauney variables as in (\ref{Hbi}). It remains to recast the interaction term $H'$ similarly, which can be replaced by $H^{(2)}$ in (\ref{Hquad}) in quadrupole approximation. To this end, for the moment we choose the reference plane to be the plane of the outer orbit, and choose the reference direction (the $x$ axis in Fig.~\ref{Fig_orbit}) to be the periapsis of the outer orbit. Then, we use $(a_i,e_i,\psi_i),~(i=1,2)$ to denote the orbital parameters of the inner and outer orbits, respectively. Since the three Euler angles $(\ga,I,\vartheta)$ define the orientation of the inner orbit relative to the outer orbit (but not the the orientations of both orbits with respect to a third reference plane), we don't assign a subscript to them. With this notation, the leading order binary motions of the two orbits can be represented by their position vectors $\mb R_{1,2}$ as,
\bge
  \mb R_1=R_1\bgp \cos\vartheta & -\sin\vartheta & 0 \\ \sin\vartheta & \cos\vartheta & 0 \\ 0 & 0 & 1 \edp
  \bgp 1 & 0 & 0 \\ 0 & \cos I & -\sin I \\ 0 & \sin I & \cos I \edp
  \bgp \cos\ga & -\sin\ga & 0 \\ \sin\ga & \cos\ga & 0 \\ 0 & 0 & 1 \edp
  \bgp \cos\psi_1 \\ \sin\psi_1 \\ 0 \edp,
\ede
and,
\bge
  \mb R_2=R_2\bgp \cos\psi_2 \\ \sin\psi_2 \\ 0 \edp,
\ede
where $R_i=|\mb R_i|=a_i(1-e_i^2)/(1+e_i\cos\psi_i),~(i=1,2)$ describe the familiar elliptical motions. Then the quadrupole Hamiltonian (\ref{Hquad}) can be written as,
\begin{align}
\label{H2}
  H^{(2)}=&-\FR{Gm_0m_1m_2}{2m}\FR{a_1^2(1-e_1^2)^2(1+e_2\cos\psi_2)^3}{a_2^3(1-e_2^2)^3(1+e_1\cos\psi_1)^2}\n\\
  &\times\Big\{3\big[\cos(\psi_1+\ga)\cos(\psi_2-\vartheta)+\sin(\psi_1+\ga)\sin(\psi_2-\vartheta)\cos I\big]^2-1\Big\}.
\end{align}

To trace the slow motion of the orbital parameters under the perturbation of $H^{(2)}$, as is standard we  ``integrate out'' the fast periodic motions of both the inner and outer orbits, which yields an ``effective Hamiltonian''. In the literature this is called the secular approximation, and the resulting effective Hamiltonian is called the doubly-averaged Hamiltonian \cite{Lidov1976}. Conventionally, this is done by averaging (\ref{H2}) over both mean anomalies $\be_{1,2}$, since mean anomalies are defined to have uniform motion in time. From (\ref{anomaly}) we know that the mean anomaly is related to the true anomaly by,
\bge
  \di\be_i=\FR{(1-e_i^2)^{3/2}}{(1+e_i\cos\psi_i)^2}\di\psi_i.~~~~(i=1,2)
\ede
Therefore, the averaged Hamiltonian $\ob{H}^{(2)}$ can be worked out as,
\begin{align}
  \ob{H}^{(2)}\equiv&~\FR{1}{4\pi^2}\int_0^{2\pi}\di\be_1\di\be_2\,H^{(2)}\n\\
  =&-\FR{Gm_0m_1m_2}{32m}\FR{a_1^2}{a_2^3(1-e_2^2)^{3/2}}\big[(2+3e_1^2)(1+3\cos2I)+30e_1^2\cos2\ga\sin^2I\big].
\end{align}
It is convenient to rewrite $\ob{H}^{(2)}$ in terms of a conserved and dimensionless function $W$ following \cite{Lidov1976}, as,
\begin{align}
  \label{H2bar}
  \ob{H}^{(2)}=&-K(W+\FR{5}{3}),\\
  \label{Kfactor}
  K\equiv&~\FR{3Gm_0m_1m_2}{8m}\FR{a_1^2}{a_2^3(1-e_2^2)^{3/2}},\\
\label{W}
  W\equiv&~(-2+\cos^2I)(1-e_1^2)+5e_1^2(\cos^2I-1)\sin^2\ga.
\end{align}

Some remarks about the perturbed Hamiltonian are as follows.
\begin{enumerate}
  \item Alternatively, it is possible to apply a canonical transformation, known as a Von Zeipel transformation, to the original Hamiltonian (\ref{Hsep}) to eliminate the fast modes corresponding to the mean anomalies $\be_{1,2}$ at the quadrupole level \cite{Naoz:2011mb}. In other words, the two mean anomalies can be eliminated without spoiling the canonical structure of the Hamiltonian. Thus we conclude that the action variables conjugate to $\be_{1,2}$, namely $J_{\be1}$ and $J_{\be2}$, are conserved quantities, and it follows immediately that the semi-major axes and energies of both orbits are separately conserved by the doubly-averaged Hamiltonian.
  
  \item The Hamiltonian $\ob{H}^{(2)}$ is obtained by substituting the leading order solutions (i.e. Keplerian orbits), and this result resembles the expectation value of the perturbed Hamiltonian calculated in the time-independent perturbation theory of quantum mechanics. In this sense we can say that (\ref{H2bar}) is an ``on-shell'' expression for $\ob{H}^{(2)}$. Consequently, while (\ref{H2bar}) can be very useful when applying ``on-shell'' arguments such as energy conservation, the (off-shell) functional dependence on canonical variables in (\ref{H2bar}) has been  obscured. Therefore, one cannot immediately derive the equations of motion from (\ref{H2bar}). In particular, it would be wrong to conclude that $J_{\vartheta1}$ and $J_{\vartheta2}$ are separately conserved from the fact that (\ref{H2bar}) is apparently independent of $\vartheta_{1,2}$. 
\end{enumerate}

To derive the equations for the secular evolution of the orbital parameters, we rewrite the Hamiltonian in terms of Delauney variables of both the inner and the outer orbits. To define these  Delauney variables, it is helpful to switch the choice of  reference plane to the one perpendicular to the total angular momentum of the triple system, known as the invariant plane. We then have six Delauney variables for each of the two orbits, but not all of them are independent dynamical variables. First, we note that  $J_{\ga i}=|\mb J_i|$ and $J_{\vartheta i}=\mb J_i\cdot \mb J/|\mb J|~(i=1,2)$, where $\mb J_{1,2}$ are angular momenta of the inner and outer orbits, and $\mb J=\mb J_1+\mb J_2$ is the total angular momentum. Therefore, these variables are related to each other via,
\begin{align}
&J_{\ga1}^2-J_{\vartheta1}^2=J_{\ga2}^2-J_{\vartheta2}^2,
&& J_{\vartheta1}+J_{\vartheta2}=J. 
\end{align}
Thus we can eliminate $J_{\vartheta1,2}$ in favor of $J_{\ga1,2}$,
\begin{align}
&J_{\vartheta1}=\FR{J^2+J_{\ga1}^2-J_{\ga2}^2}{2J},
&&J_{\vartheta2}=\FR{J^2-J_{\ga1}^2+J_{\ga2}^2}{2J}.
\end{align}
Note that the quadrupole Hamiltonian (\ref{H2bar}) is independent of $\vartheta_{1,2}$, and thus we have removed the conjugate pair $(J_{\vartheta,i},\vartheta_i)$. Secondly, we have noted before that $J_{\be1,2}$ are conserved quantities. Finally, one can see that $\ga_1=\ga$ where $\ga$ is defined above with respect to the outer orbital plane, since the angular momenta of the two orbits are coplanar with the total angular momentum. Therefore, we see that the quadrupole Hamiltonian (\ref{H2bar}) is also independent of $\ga_2$. These arguments reduce the triple system into an integrable system with only one pair of conjugate variables $\ga_1$ and $J_{\ga1}$. We further note that the total inclination $I=I_1+I_2$, i.e. the inclination of the inner orbit relative to the outer orbit is related to various angular momenta via,
\bge
\label{cosI}
   \cos I=\FR{J^2-J_{\ga1}^2-J_{\ga2}^2}{2J_{\ga 1}J_{\ga 2}}.
\ede
 Now we can rewrite the quadratic Hamiltonian (\ref{H2bar}) in terms of $\ga_1$ and $J_{\ga1,2}$, as,
\begin{align}
  \ob{H}^{(2)}=&-\FR{3G\mu_1m_2}{8}\FR{a_1^2J_{\be2}^3}{a_2^3J_{\ga2}^3}\bigg\{\FR{J_{\ga1}^2}{J_{\be1}^2}\bigg[-2+\Big(\FR{J_{\ga1}^2+J_{\ga2}^2-J^2}{2J_{\ga1}J_{\ga2}}\Big)^2\bigg]\n\\
  &+5\bigg(1-\FR{J_{\ga1}^2}{J_{\be1}^2}\bigg)\bigg[\Big(\FR{J_{\ga1}^2+J_{\ga2}^2-J^2}{2J_{\ga1}J_{\ga2}}\Big)^2-1\bigg]\sin^2\ga_1+\FR{5}{3}\bigg\}.
\end{align}
Then the equations for the orbital parameters can be derived from the canonical equations $\dot\ga_1=\pd \ob{H}^{(2)}/\pd J_{\ga1}$ and $\dot J_{\ga1}=-\pd \ob{H}^{(2)}/\pd \ga_1$.
\begin{align}
\label{dedtKL}
\FR{\di e_1}{\di t}\bigg|_\text{KL}=&~\FR{5K}{J_{\ga1}}e_1(1-e_1^2)(1-\cos^2I)\sin 2\ga_1,\\
\label{dgdtKL}
\FR{\di \ga_1}{\di t}\bigg|_\text{KL}=&~2K\bigg[\FR{1}{J_{\ga1}}\big(2(1-e_1^2)-5(1-e_1^2-\cos^2I)\sin^2\ga_1\big) \n\\&~~~~~~+\FR{1}{J_{\ga2}}\big(1-e_1^2+5e_1^2\sin^2\ga_1\big)\cos I\bigg].
\end{align}
We now have the equations of motion governing the secular evolution of  the inner binary. When applied to inspiraling BBHs, they should be supplemented by the equations from the PN correction and GW back reactions, which we elaborate in the following.

\subsubsection{Post-Newtonian correction}

As is well known, the first nontrivial order of the PN correction to the Keplerian potential is a trivial constant shift plus a new term proportional to the inverse square of the distance, for which the net effect is to generate a precession for the periapsis of an elliptical orbit. The precession rate is given by (See, e.g., \cite{Weinberg:1972kfs}),
\bge
\label{dgammadtPN}
  \FR{\di\ga_1}{\di t}\bigg|_\text{PN}=\FR{3}{c^2a_1(1-e_1^2)}\bigg(\FR{Gm}{a_1}\bigg)^{3/2}.
\ede
While we can just include this term in the evolution equation for $\ga_1$ and study its effect by solving the equation, it will be useful to reconstruct a corresponding term in the Hamiltonian, which will allow us to apply an energy conservation argument to estimate the merger time in (\ref{emax}). Since a $1/r^2$ potential results from a conservative force, the corresponding  Hamiltonian term must also be conserved. In principle one can derive the desired Hamiltonian again from the original PN Hamiltonian by double averaging over $\be_{1,2}$. Here we adopt a simpler poor man's derivation by rewriting $\dot\ga_1|_\text{PN}$ in terms of Delauney's variables and integrating,
\begin{align}
  \FR{\pd H_\text{PN}}{\pd J_{\ga1}}=\FR{\di\ga_1}{\di t}\bigg|_\text{PN}=\FR{3}{c^2}\FR{Gm\mu^2}{J_{\ga1}^2}\bigg(\FR{Gm\mu}{J_{\be1}}\bigg)^3~~\To~~H_\text{PN}=\int\di J_{\ga1}\FR{\di\ga_1}{\di t}\bigg|_\text{PN}=-\FR{3G^2m^2\mu}{c^2a_1^2\sqrt{1-e_1^2}}.
\end{align}
It is convenient to rewrite $H_\text{PN}$ as,
\begin{align}
\label{HPN}
  &H_\text{PN}=-KW_\text{PN},\\
\label{WPN}
  &W_\text{PN}=\FR{8Gm^2a_2^3(1-e_2^2)^{3/2}}{c^2m_2a_1^4(1-e_1^2)^{1/2}}.
\end{align}
where $K$ is defined in (\ref{Kfactor}). In the following, we shall also use the parameter $\Theta_\text{PN}\equiv W_\text{PN}\sqrt{1-e_1^2}$ which is conventional in the literature. We note that (\ref{HPN}), as with  (\ref{H2bar}), is  an ``on-shell'' expression that can be used when applying energy conservation, but  is not suitable for deriving the equation of motion. 

  There are additional PN corrections which we have not included.  The correction to the quadrupole coupling between the orbits is subdominant in that it is smaller than the Newtonian quadrupole which is already a perturbation. The PN correction to the outer orbit gives nonzero $\dot \ga_2$, which is irrelevant to the KL oscillations at quadrupole order since it is independent of $\ga_2$.  This is no longer valid in systems such as globular clusters when octupole interactions play an essential role \cite{Naoz:2012bx}. An $N$-body code of the triple system with PN corrections were presented in \cite{Bonetti:2016eif} and further subtleties in the calculation of GW radiation from a hierarchical triple were discussed in \cite{Bonetti:2017hnb}. 

\subsubsection{Gravitational wave radiation}

The radiated GWs from the inner binary carry energy and angular momentum away from the system, leading to a reduction of both the  semi-major axis $a$ and the eccentricity $e$. Throughout the paper we assume that the semi-major axis of the outer orbit is much greater than that of the inner orbit so that the GW back reaction on the outer orbit can be neglected and we can consider the influence of GWs only on $a_1$ and $e_1$. The equations governing $\dot a_1$ and $\dot e_1$, known as Peters's equations \cite{Peters:1964zz}, have been reviewed in \cite{Randall:2017jop} and we only quote the results here.
\begin{align}
\label{dadt}
\FR{\di a_1}{\di t}\bigg|_\text{GW}=&-\FR{64}{5}\FR{G^3\mu_1 m^2}{c^5a_1^3}\FR{1}{(1-e_1^2)^{7/2}}\bigg(1+\FR{73}{24}e_1^2+\FR{37}{96}e_1^4\bigg),\\
\label{dedt}
\FR{\di e_1}{\di t}\bigg|_\text{GW}=&-\FR{304}{15}\FR{G^3\mu_1 m^2}{c^5a_1^4}\FR{e_1}{(1-e_1^2)^{5/2}}\bigg(1+\FR{121}{304}e_1^2\bigg).
\end{align}
We solve $a_1$ as a function of $e_1$ from the above two equations as $a_1=g(e_1)$ with $g(e)$ defined as,
\bge
\label{ge}
  g(e)=\FR{e^{12/19}}{1-e^2}\bigg(1+\FR{121}{304}e^2\bigg)^{870/2299}\simeq\left\{
  \begin{split}
    &e^{12/19},  &(e\ll 1)\\
    &\FR{1.1352}{1-e^2}. & (e\lesssim 1)
  \end{split}
  \right.
\ede

The GW radiation affects the evolution of $a_1$ and $e_1$ not only through Peters's equation above, but also implicitly through (\ref{dedtKL}) and (\ref{dgdtKL}) since the right hand side of these two equations depends on the inclination, which in turn depends on the total angular momentum $J$ through (\ref{cosI}). Since the total angular momentum is no longer conserved after including GWs, we should add another equation describing the loss of it. Remember that we consider the GW back reaction to  the inner orbit only, and thus the loss of angular momentum happens only  to $|\mb J_1|=J_{\ga 1}$ but not to $|\mb J_2|$. This loss of angular momentum has been reviewed in \cite{Randall:2017jop} and can be written as,
\bge
  \FR{\di J_{\ga 1}}{\di t}=-\FR{32}{5}\FR{G^{7/2}\mu_1^2 m^{5/2}}{c^5a_1^{7/2}}\FR{1}{(1-e_1^2)^2}\bigg(1+\FR{7}{8}e_1^2\bigg).
\ede
Then differentiating (\ref{cosI}), we get $\dot J=(J_{\ga 1}+J_{\ga 2}\cos I)\dot J_{\ga 1}/J$, where we have used the fact that the GW doesn't affect the inclination $I$ since it only reduces the magnitude of the angular momentum but does not influence its direction. Consequently,
\bge
\label{dJdt}
  \FR{\di J}{\di t}\bigg|_\text{GW}=-\FR{32}{5}\FR{G^{7/2}\mu_1^2 m^{5/2}}{c^5a_1^{7/2}}\FR{J_{\ga 1}+J_{\ga 2}\cos I}{J}\FR{1}{(1-e_1^2)^2}\bigg(1+\FR{7}{8}e_1^2\bigg).
\ede

\subsubsection{Summary}

All three pieces --- the quadrupole interaction between the inner and outer orbits, the PN precession of the inner orbit, and the GW back reaction --- can now be assembled to give the following set of equations.
\bgs
\label{secularEq}
\begin{align}
\FR{\di a_1}{\di t}=&-\FR{64}{5}\FR{G^3\mu_1 m^2}{c^5a_1^3}\FR{1}{(1-e_1^2)^{7/2}}\bigg(1+\FR{73}{24}e_1^2+\FR{37}{96}e_1^4\bigg),\\
\FR{\di e_1}{\di t}=&~\FR{5K}{J_{\ga1}}e_1(1-e_1^2)(1-\cos^2I)\sin 2\ga_1-\FR{304}{15}\FR{G^3\mu_1 m^2}{c^5a_1^4}\FR{e_1}{(1-e_1^2)^{5/2}}\bigg(1+\FR{121}{304}e_1^2\bigg),\\
\FR{\di \ga_1}{\di t}=&~2K\bigg[\FR{1}{J_{\ga1}}\big(2(1-e_1^2)-5(1-e_1^2-\cos^2I)\sin^2\ga_1\big) \n\\&~~~~~~+\FR{1}{J_{\ga2}}\big(1-e_1^2+5e_1^2\sin^2\ga_1\big)\cos I\bigg]+\FR{3}{c^2a_1(1-e_1^2)}\bigg(\FR{Gm}{a_1}\bigg)^{3/2},\\
\FR{\di J}{\di t}=&-\FR{32}{5}\FR{G^{7/2}\mu_1^2 m^{5/2}}{c^5a_1^{7/2}}\FR{J_{\ga 1}+J_{\ga 2}\cos I}{J}\FR{1}{(1-e_1^2)^2}\bigg(1+\FR{7}{8}e_1^2\bigg).
\end{align}
\eds
This set of equations are to be supplemented by the definition of $K$ in (\ref{Kfactor}), the formulae for $J_{\ga1,2}$ in (\ref{actionvariable}), the relation (\ref{cosI}) between the inclination $I$ and the total angular momenta $J$. Then, everything can be expressed in terms of the four independent variables $(a_1,e_1,\ga_1,J)$.

At this point, in principle we can sample the initial parameters following a given distribution, and solve this set of equations to get the final distribution of eccentricity at the LIGO threshold. But it is far preferable to find a direct analytical relation between the  initial parameters and the final eccentricity. Such a solution would presumably elucidate  the dependence of the final answer on the  input parameters, and the calculation would also be much faster than directly solving the equations case by case -- hopefully yielding some insight into the environments in which observation binary black holes mergers occur. The challenge of course is the large number of parameters, but an analytical solution will help pinpoint what might be possible to extract. In the next section we set out to find this analytical mapping between initial parameters and final eccentricity. We can then get the final eccentricity distribution by integrating over the initial parameter space, given any initial distribution of binaries.

This problem involves a large number of parameters, but some should be measurable by LIGO and only several of the others significantly impact the resultant eccentricity. In principle this makes it feasible to learn about the environment of the merger. Specifically, the solutions depend on the following parameters: the three masses $(m_{0},m_1,m_2)$, two constant parameters of the outer orbit $(a_2,e_2)$, and four initial conditions, which we can choose to be $(a_1,e_1,\ga_1,I)$. Here we replace the magnitude of the total angular momentum $J$ in (\ref{secularEq}) by the inclination $I$ by means of (\ref{cosI}). Given these parameters, we can follow the evolution of the triple system until the inner binary separation $a_1$ reaches the observational band. Among these parameters, two masses $(m_0,m_1)$ together with the inner orbital parameter $(a_1,e_1)$ are measurable after entering the LIGO band. The other parameters are not directly accessible from LIGO observation, but they can make an impact on the observables $e_1$. In the next section we will see that for slow mergers $(a_2,e_2)$ and $I$ have the strongest impact on the final $e_1$ while the impact of $(a_{1},e_{1},\ga_{1})$ is relatively mild.

\section{Merger of the Inner Binary}
\label{sec_merger}

In this section we develop an analytical understanding of the merger of the inner binary. In order to provide some intuition, we first present several typical solutions to the secular equations in Sec.\;\ref{sec_NumExp}. We will see that according to the initial conditions,  the inner binary can undergo three qualitatively distinct modes of evolution.

We estimate the merger time and the eccentricity at the LIGO threshold of the inner binary in each of the three cases in Sec.\;\ref{sec_MergerTime}. 

We complete the analysis begun in our previous paper \cite{Randall:2017jop} by showing that the final eccentricity at the LIGO threshold can be estimated from the merger time. The distribution of eccentricity $f(e)$ at the beginning of GW domination, though calculable, is no longer needed.

\subsection{Numerical Examples}
\label{sec_NumExp}

In this subsection we present several qualitatively distinct solutions for the inner binary merger, all with initially highly inclined orbits. As we explore below, the difference in relative strengths of the KL oscillation and the post-Newtonian precession is critical to the merger as the former increases eccentricity while the latter suppresses it. The type of numerical solutions we present are not new but we review them to gain some intuition that will be helpful when constructing our analytical method. Our numerical samples are generated using the secular equations (\ref{secularEq}), which are free of numerical subtleties that could occur in a direct $N$-body simulation. A nice agreement between integrating secular equations and the direct $N$-body simulation has been shown in \cite{Antonini:2012ad} using  the \verb+ARCHAIN+ integrator. Similar numerical solutions have been presented in \cite{Wen:2002km} but for different parameters.

Throughout this subsection, we set the binary masses $m_0=m_1=10M_\odot$, the tertiary mass $m_2=4\times 10^6M_\odot$. We assume that the initial inner semi-major axis is $a_{10}=0.1$AU, and the initial inner and outer eccentricities are $e_{10}=e_{20}=0.1$. Furthermore, we choose the initial inclination to be $I_0=89.9^\circ$ and the initial argument of inner periapsis $\ga_{10}=0^\circ$. By changing the initial value of the outer semi-major axis $a_{20}$, we can adjust the strength of the KL oscillation relative to PN corrections. 

\begin{figure}[tbph]
\centering
\includegraphics[height=53mm]{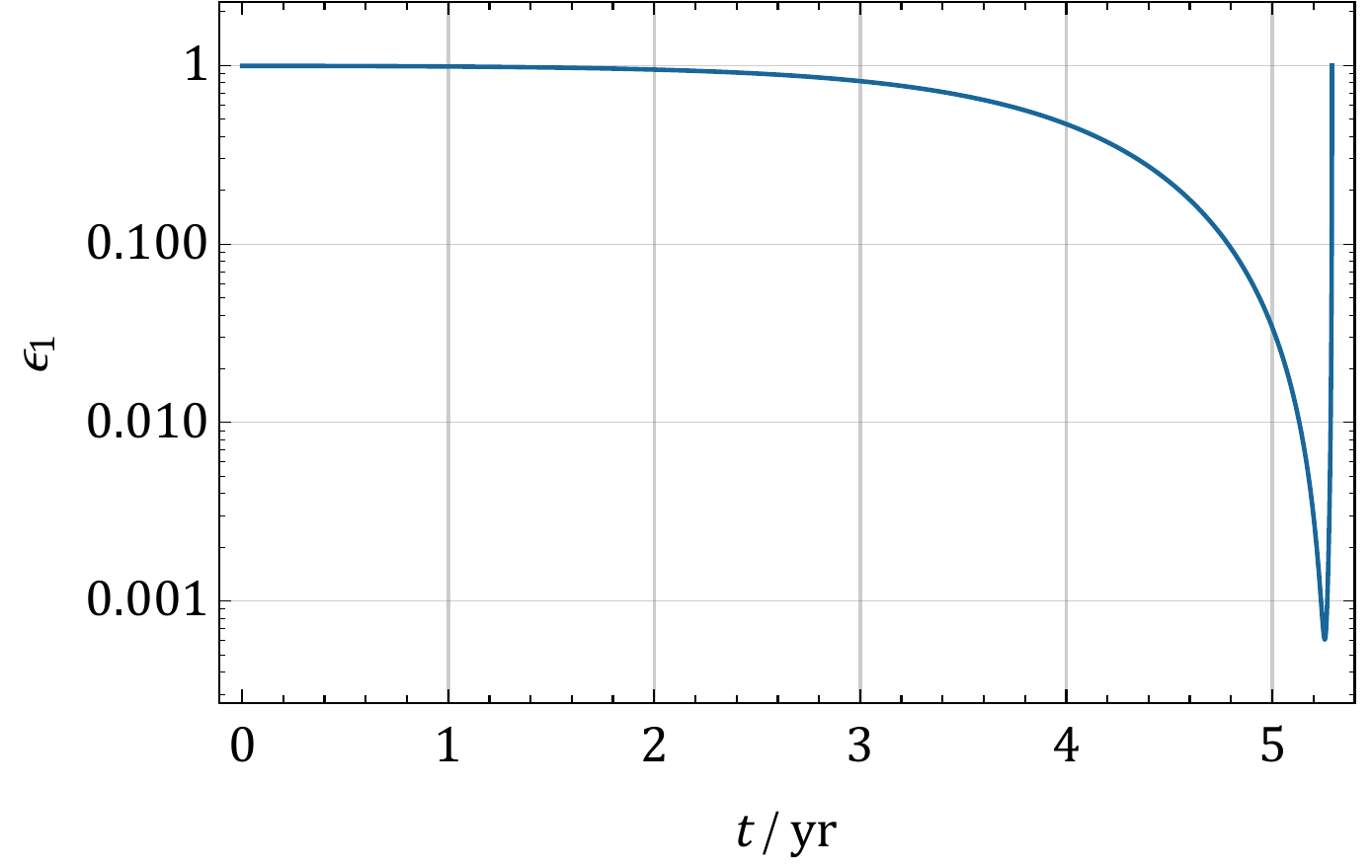}\hspace{3mm}
\includegraphics[height=53mm]{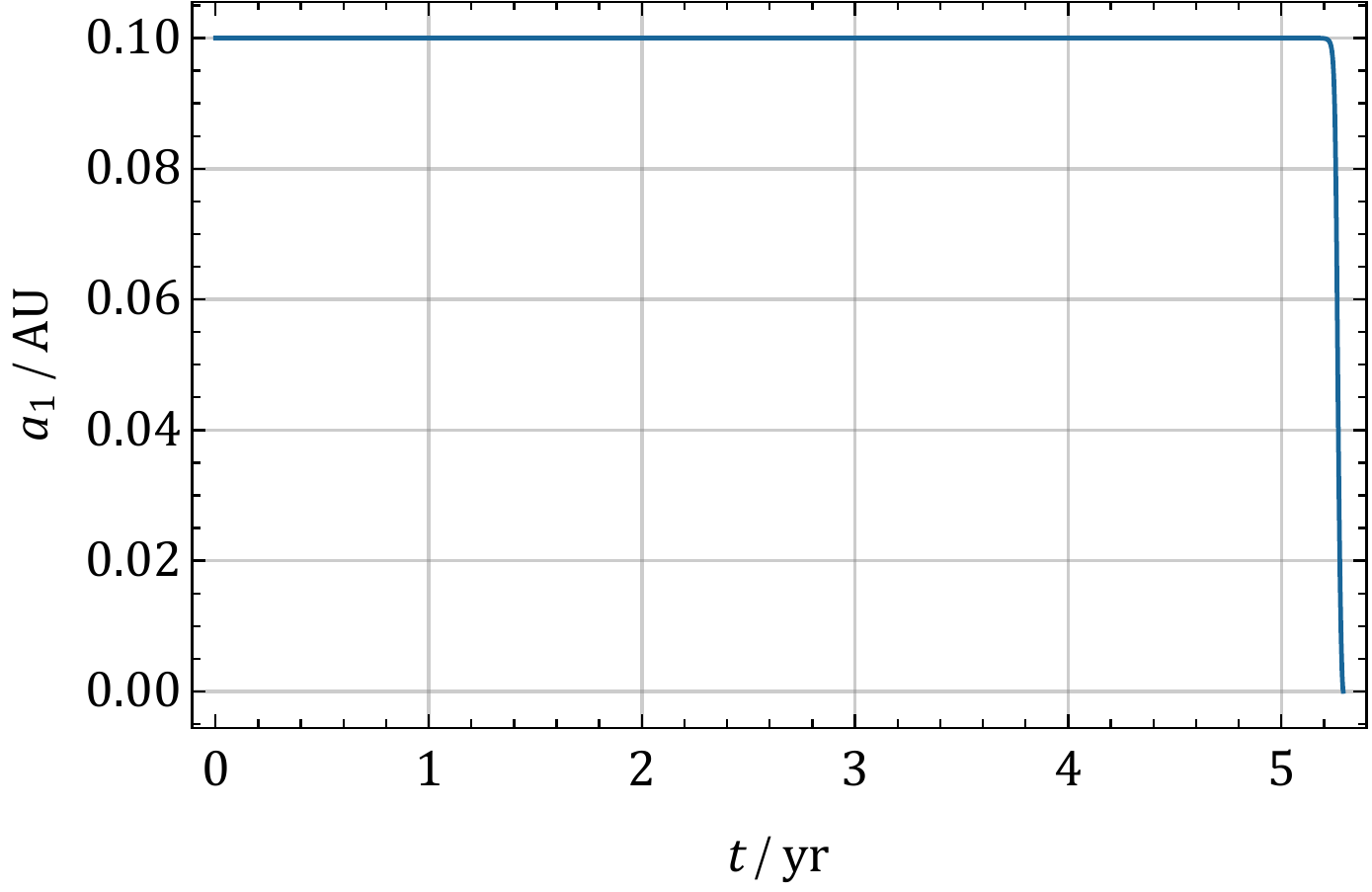}
\caption{$\ep_1(t)$ and $a_1(t)$ in Example 1, with $a_{20}=80$AU.}
\label{Fig_Sample1}
\end{figure}

For our first example, we take $a_{20}=80$AU, which means that the inner binary BHs are close enough to the central SMBH  that the KL resonance is extremely effective. The maximal value of the eccentricity in this case can be so large that the binary merges within one KL cycle. As we can see from Fig.\;\ref{Fig_Sample1}, the merger happens when $\ep_1=1-e_1^2$ reaches its minimum $\ep_{1\text{min}}\sim\order{0.001}$ on the time scale of the KL oscillation which is about $\simeq 5$yrs in this example. 
\begin{figure}[tbph]
\centering
\includegraphics[height=56mm]{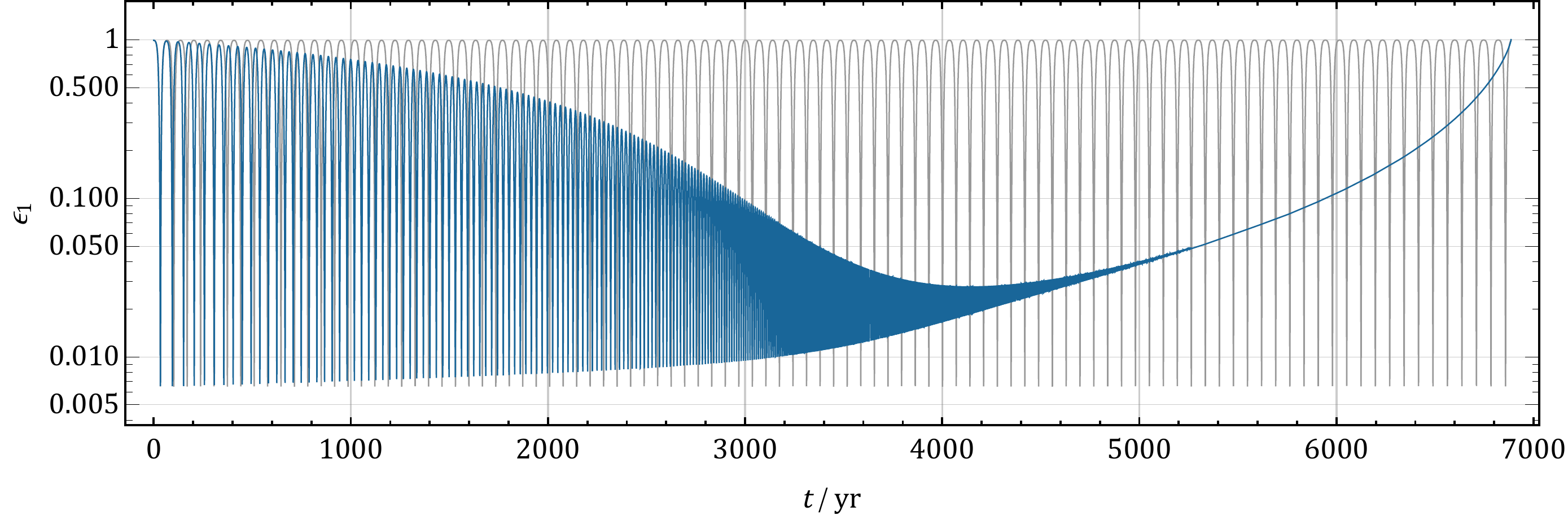}\\
\includegraphics[height=50mm]{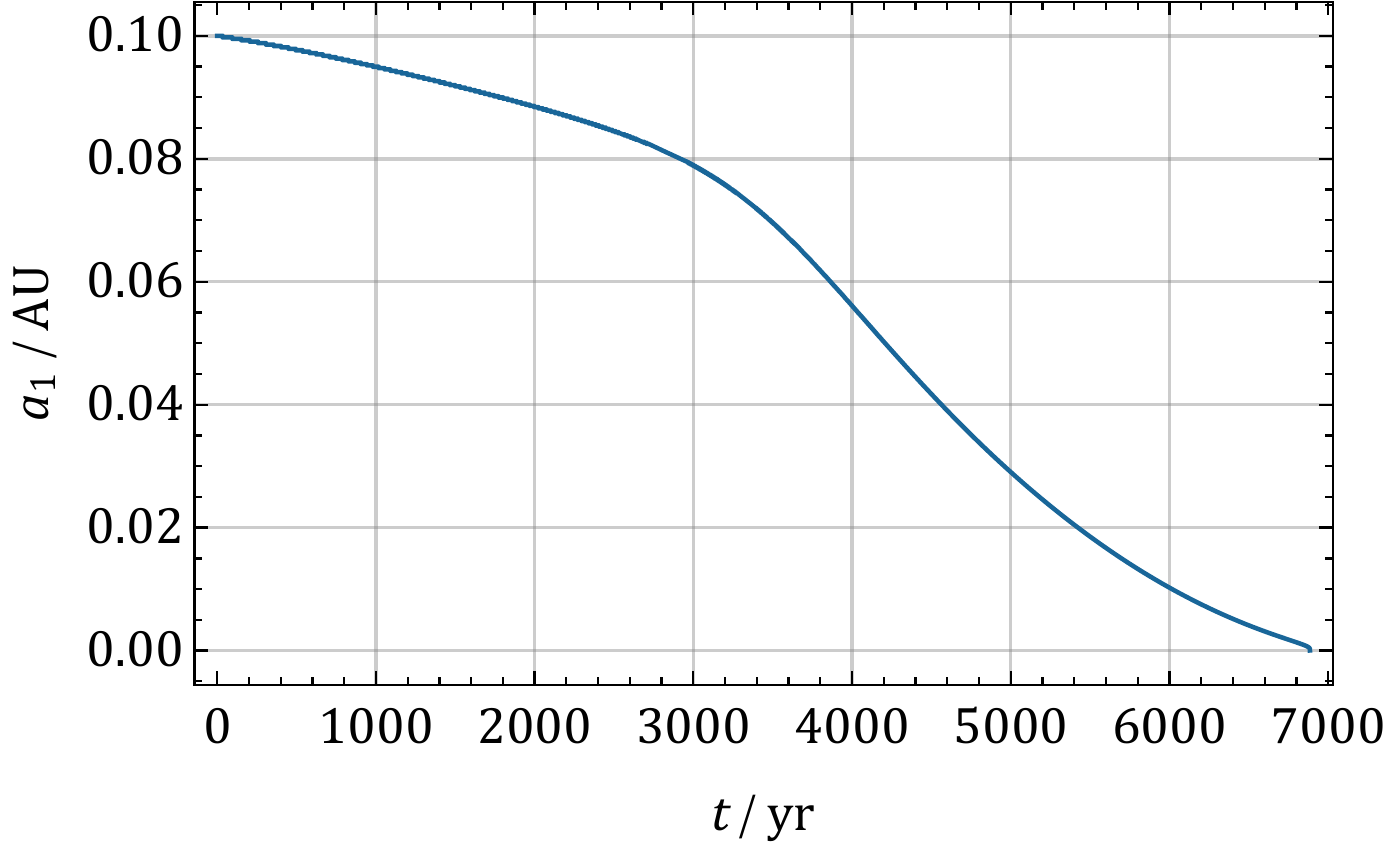}\hspace{3mm}
\includegraphics[height=50mm]{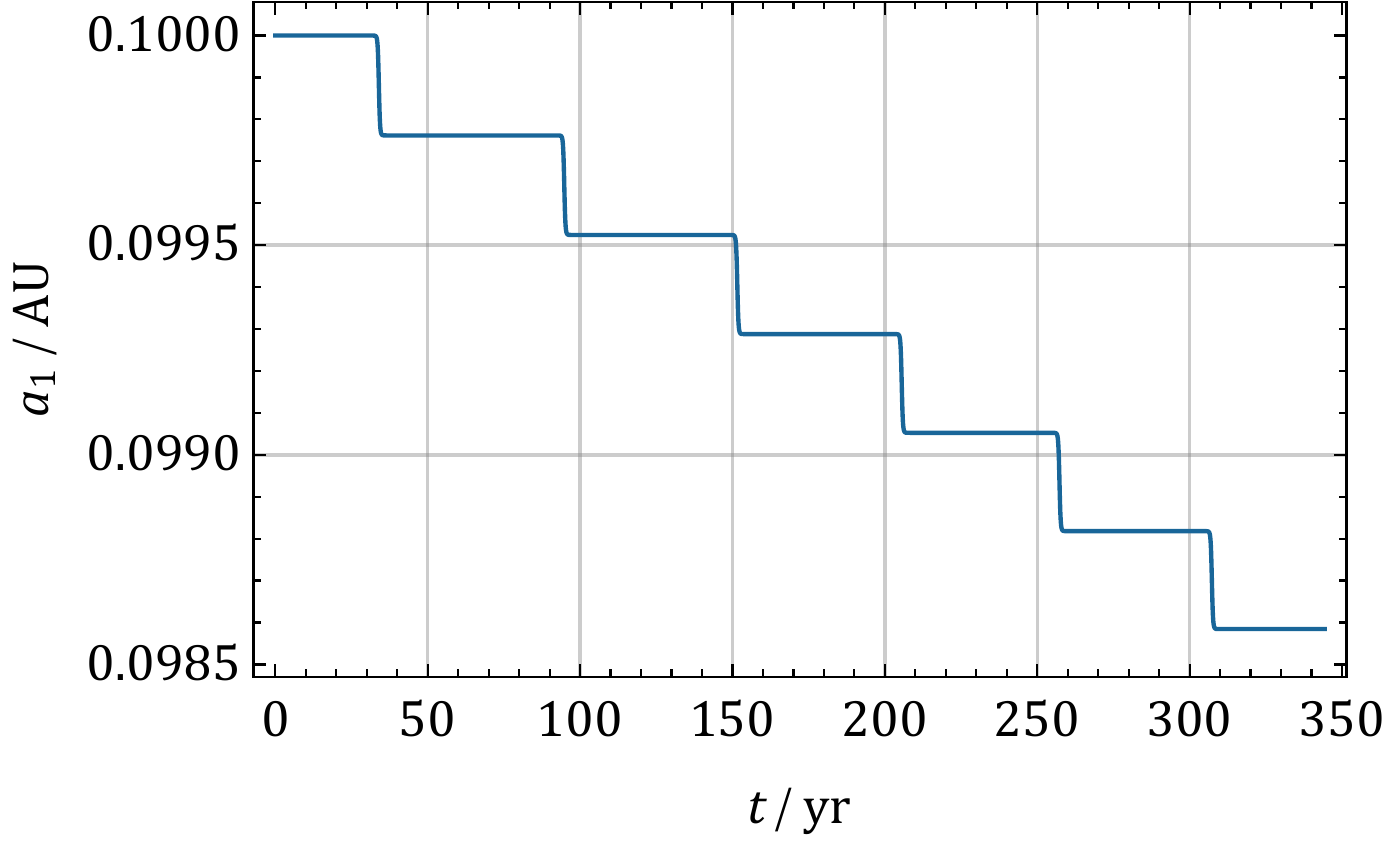}
\caption{$\ep_1(t)$ and $a_1(t)$ in Example 2, with $a_{20}=150$AU. Upper panel: the solutions of $\ep_1(t)$ from the full evolution equations (blue curve) and from the equations without GWs (gray curve). Lower-left panel: $a_1(t)$; Lower-right panel: $a_1(t)$ zoomed in.}
\label{Fig_Sample2}
\end{figure}

In our second example, we take $a_{20}=150$AU, which is about twice the distance of the first example. The KL time scale is correspondingly longer. In initial several KL cycles, the KL time scale $\sim(K/J_{\ga 1})^{-1}$ is about fifty years, whereas the PN time scale $(\di\ga_1/\di t|_\text{PN})^{-1}$ is about one year at $\ep_{1\text{min}}$, as can be found from (\ref{dgammadtPN}).  So the PN precession is more effective in suppressing the maximal value of $e_1$ than it in the first example.  As a result, the reduction of $a_1$ in each KL cycle is less significant, and the merger of the inner binary takes a large number of cycles.

From Fig.\;\ref{Fig_Sample2} we see that the merger time in this example is $\sim 7000$yrs. In its early stage, $\ep_1$ can reach a minimum $\sim 0.006$. This minimum doesn't change significantly in the first half of the binary's life, but it becomes larger in the later stage due to the stronger effect of  GW emission. Were there no GWs, the frequency of the oscillation and the value of $\ep_{1\text{min}}$ would stay constant as one can see from the grey curve in the top panel of Fig.\;\ref{Fig_Sample2}, since PN precession conserves energy and angular momentum. Accordingly, the semi-major axis $a_1(t)$ changes more slowly than in the first example. However, if we zoom in  to examine the function $a_1(t)$ on  smaller time scales, we see that, at least during the early stage of the merger, the reduction of $a_1$ occurs chiefly at the minimum of $\ep_1$, so that $a_1(t)$ has a stair-wise behavior. In addition, it can be seen from Fig.\;\ref{Fig_Sample2} that the period of the KL oscillation decreases over time, and that the minimal/maximal $e_1$ (i.e. maximal/minimal $\ep_1$) in each KL cycle increases/decreases monotonically with time until the KL oscillation is almost fully suppressed. All these phenomena can be understood by noting that the period of the KL oscillation in $e_1$ is effectively determined by $\ga_1$ as one can see from (\ref{dedtKL}). The PN correction always advances the phase of $\ga_1$ relative to the phases in its absence, and thus effectively reduces the period and amplitude of the KL oscillation and hence eccentricity. The strength of PN precession, when expressed in terms of $W_\text{PN}$ in (\ref{WPN}), is proportional to $a_1^{-4}$. Therefore the PN correction gets stronger at later stage when $a_1$ is smaller. Consequently, the amplitude of the KL oscillation which is in any case smaller with smaller $a_1$, is even smaller at later time.

\begin{figure}[tbph]
\centering
\includegraphics[height=53mm]{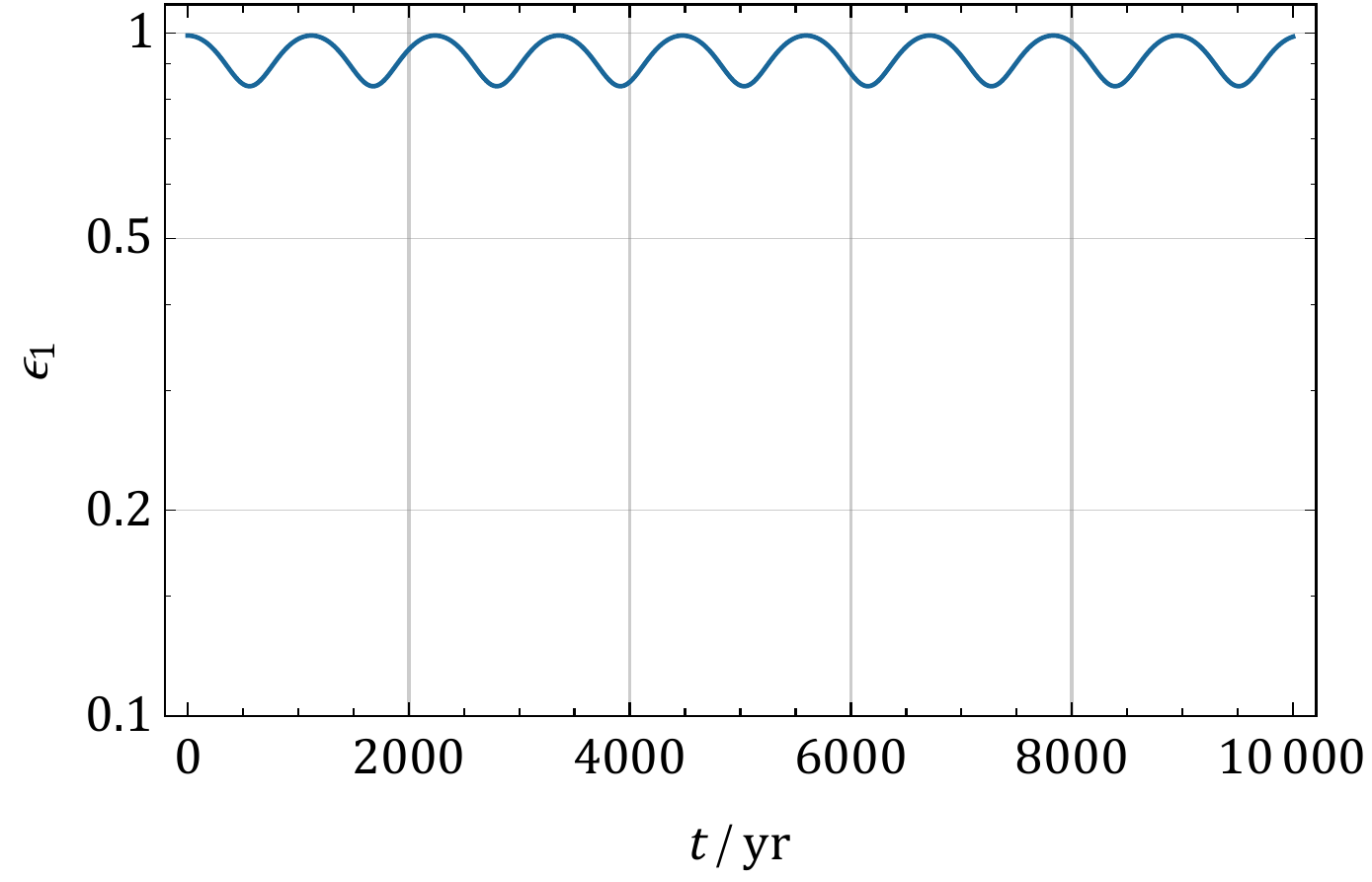}\hspace{3mm}
\includegraphics[height=53mm]{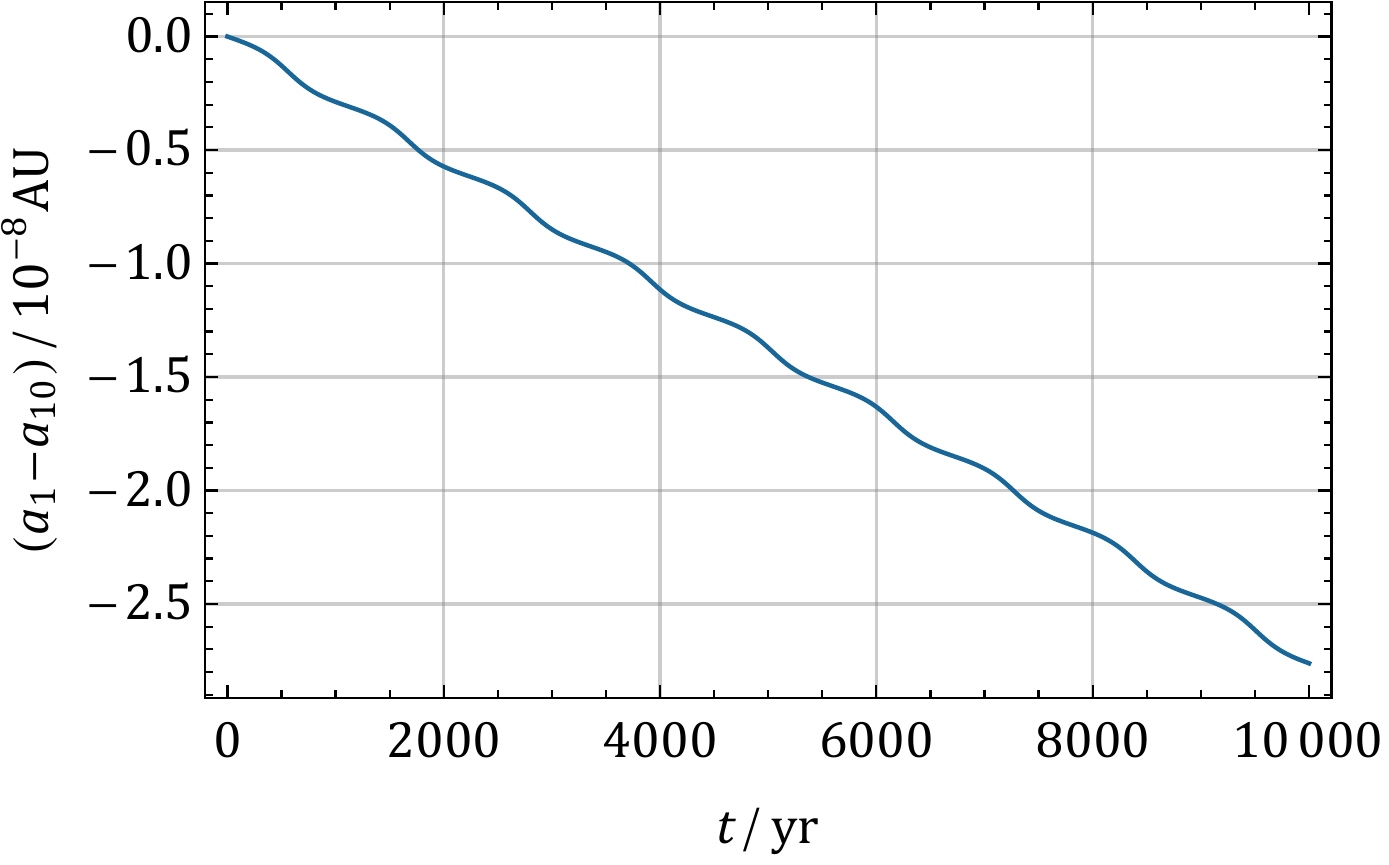}
\caption{$\ep_1(t)$ and $a_1(t)$ in Example 3, with $a_{20}=420$AU.}
\label{Fig_Sample3}
\end{figure}
For our third example, we choose $a_2=420$AU. In this case, the PN precession has such a strong effect on the KL cycle and the maximum of $e_1$ (thus the minimum of $\ep_1$) that the KL oscillation is strongly suppressed. From Fig.\;\ref{Fig_Sample3} we see that $\ep_1$ in this case never becomes small enough to boost GW radiation. Consequently, the stair-wise function $a_1(t)$ in the second example gets ``melted'' into a smoother function of time. Therefore, the merger time in such cases  is well approximated by the merger time of an isolated binary with initial eccentricity given by $e_{1\text{max}}$, where $e_{1\text{max}}$ is the maximal value of eccentricity during its tiny oscillations, and can be extremely long. 

\subsection{Analytic Estimate of the Merger Time}
\label{sec_MergerTime}

Now we estimate the merger time $\tau$ of the inner binary, which is defined to be the time the binary takes to coalesce with given initial parameters.  Recall that this parameter is critical to determining if a binary will merge or evaporate. As the previous numerical solutions show, there are three distinct regions of parameter space depending on the maximal value of eccentricity that the inner binary can reach: 1) For highly inclined binary orbits that are very close to a central BH, the eccentricity can be close to one, in which case the merger happens within one KL cycle. We call this scenario a \emph{fast merger}. 2) For the intermediate case,  which we call \emph{KL-boosted}, the merger happens after many KL cycles, and the reduction of $a_1$ happens chiefly at those times within a cycle when the eccentricity reaches the maximum. 3) For binaries very far from the central BH, no effective KL cycles exist and the reduction of $a_1$ is essentially a smooth function of $t$, and we call this case the \emph{isolation limit}. We call both KL-boosted binaries and binaries in the isolation limit \emph{slow mergers}. 

Due to the qualitatively different behavior in the three cases, we need to estimate the merger time for each of them independently. We assume arbitrary initial values of the  triple parameters, including $a_{1,2}$, $e_{1,2}$, and $I$. However, we assume $\ga_1=0$ because we can always let the inner binary evolve to $\ga_1=0$ were it not so initially. Clearly this assumption breaks down for the fast merger because the binary can merge before $\ga_{1}$ evolves to 0. However, even in this case it is still sensible to assume $\ga_{10}=0$ since the merger time estimated in this way is correct within an order-of-magnitude. Furthermore, the initial value of $e_{10}$ will change if we shift the initial time to $\ga=0$. Therefore assuming $\ga_{1}=0$ initially will affect the initial distribution of $e_{10}$. But we will show later on that the final eccentricity at the LIGO threshold is insensitive to the initial distribution of $e_{10}$, so this correction is unimportant when integrating over initial distributions. 
 
Now we estimate the merger time for the three cases, firstly the isolation limit, then the KL-boosted case, and finally the fast mergers.

\paragraph{Isolation limit.} When tidal forces are small, the merger time $\tau_\text{iso}$ can be well approximated by the merger time of an isolated binary of the same initial $a_{10}$ and $e_{10}$. Therefore we can get $\tau_\text{iso}$ by integrating (\ref{dadt}) from $a_1=a_{10}$ to the coalescence $a_1=0$. Practically it is more convenient to integrate (\ref{dedt}) instead, from $e_1=e_{10}$ to $e_1=0$, since we know from (\ref{ge}) that $e_1$ is small at small $a_1$ at  coalescence. See \cite{MaggioreGW} for more details. The resulting $\tau_\text{iso}$ is,
\begin{align}
\label{tiso}
  \tau_\text{iso}=\FR{5}{256}\FR{c^5a_{10}^4}{G^3m_0m_1m}G(e_{10})(1-e_{10}^2)^{7/2},
\end{align}
with $G(e_{10})$ defined by the following integral,
\begin{align}
  G(e_{10})=\FR{48}{19g^4(e_{10})(1-e_{10}^2)^{7/2}}\int_0^{e_{10}}\di e\,\FR{g^4(e)(1-e^2)^{5/2}}{e(1+\frac{121}{304}e^2)}.
\end{align}
Using $g(e)\sim e^{19/12}$ for small $e$ in (\ref{ge}), we can see that the function $G(e_{10})$ equals $1$ when $e_{10}=0$, and increases monotonically with $e_{10}$ but remains very close to 1 for most of $e_{10}\in(0,1)$, until it rapidly rises to $768/425\simeq 1.80$ when $e_{10}\to 1$. Therefore it is a good approximation to neglect $G_{10}$ and just use the following expression for our estimate,
\begin{align}
\label{tiso}
  \tau_\text{iso}\simeq \FR{5}{256}\FR{c^5a_{10}^4}{G^3m_0m_1m}(1-e_{10}^2)^{7/2}.
\end{align}

\paragraph{KL-boosted.} For KL-boosted mergers, the reduction of $a_1$ happens mostly when $\ep_1$ is around its minimum (i.e. the eccentricity $e_1$ reaches its maximum) during each KL cycle, since we learn from (\ref{dadt}) that $\dot a_1\propto \ep_1^{-7/2}$. More explicitly, when $\ep_1$ increases from its minimum $\ep_{\text{1min}}$ by a factor of 2, $\dot a_1\propto $ drops to $ 9\%$ of its maximal value at $\ep_{1\text{min}}$.  We now show how to estimate the merger time by considering an imagined and isolated binary, with initial separation given by $a_{10}$ and initial eccentricity set to $e_\text{1max}$.  Here $e_\text{1max}$ is taken to be the maximal value of $e_1$ in the first several KL cycles, where $e_\text{1max}$ does not change significantly. To proceed along these lines we need to determine the maximal eccentricity of the first KL cycle $e_\text{1max}$, or equivalently, $\ep_\text{1min}\equiv 1-e_\text{1max}^2$.

For those mergers where the GW back reaction is negligible in the first several KL cycles, the value $\ep_\text{1min}$ can be estimated by the conservation of energy, that is, the sum $\ob H^{(2)}+H_\text{PN}$, or more conveniently $W+W_\text{PN}$, is a constant. Here we have three time-dependent parameters $e_1$, $I$, and $\ga_1$. We input the initial values $e_\text{10}$ and $I_0$, while, as observed earlier, the initial value of $\ga_{1}$ can always be set to zero without loss of generality because otherwise we can just wait until $\ga_1$ returns to 0. On the other hand, we see from either (\ref{W}) or (\ref{dedtKL}) that the eccentricity $e_1$ reaches its maximum when $\ga_1=\pi/2$. Therefore, we can solve $e_\text{1max}$ from the following equation,
\bge
\label{emax}
  \Big[W+W_{\text{PN}}\Big]_{e_1=e_{10},I=I_0,\ga_1=0}=\Big[W+W_{\text{PN}}\Big]_{e_1=e_\text{1max},\ga_1=\pi/2}.
\ede
It is possible to derive an analytical approximation for $\ep_\text{1min}$ when $\ep_\text{1min}\ll 1$ \cite{Wen:2002km}. To see this, we spell out the equation (\ref{emax}) explicitly as follows, 
\begin{align}
\label{emax2}
 (-2+\cos^2 I_0)\ep_{10}+\FR{\Theta_\text{PN}}{\sqrt{\ep_{10}}}=(-2+\cos^2 I_\star)\ep_\text{1min}+5(1-\ep_\text{1min})(\cos^2 I_\star -1)+\FR{\Theta_\text{PN}}{\sqrt{\ep_\text{1min}}},
\end{align}
where $\Theta_\text{PN}=W_\text{PN}\sqrt{1-e_1^2}$ as we defined earlier below (\ref{WPN}),  and $I_\star$ is the inclination $I$ evaluated at $e_1=e_{1\text{max}}$ and $\ga_1=\pi/2$. The value of $I_\star$ is not independent and can be determined by the conservation of angular momentum.  To see this, we rewrite (\ref{cosI}) as,
\bge
  J_{\ga 1}\bigg(\cos I+\FR{J_{\ga1}}{2J_{\ga 2}}\bigg)=\FR{J^2-J_{\ga2}^2}{2J_{\ga 2}}.
\ede
Both $J$ and $J_{\ga2}$ are conserved quantities. therefore the combination on the right-hand side (and hence left-hand side) of the above formula must be conserved as well. We evaluate this combination with both $(\ep_{10},I_0,\ga_{10}=0)$ and $(\ep_\text{1min},I_\star,\ga_1=\pi/2)$, and equate them, so that $I_\star$ can be solved to be,
\bge
  \cos^2 I_\star\simeq \FR{\ep_{10}}{\ep_\text{1min}}\bigg(\cos I_0+\FR{\sqrt{\ep_{10}}}{J_{\ga2}/J_{\be1}}\bigg)^2,
\ede
where we have used the condition $\sqrt{\ep_\text{1min}}/(J_{\ga2}/J_{\be1})\ll \cos I_\star$,  which is true because both factors on the left-hand side are much smaller than one while the right-hand side is of $\order{1}$. Substituting this solution back into (\ref{emax2}), and keeping the leading terms in $\ep_\text{1min}\ll 1$ limit, we get a quadratic equation for $\sqrt{\ep_\text{1min}}$ which can be readily solved as,
\begin{align}
\label{eminsol}
  \sqrt{\ep_\text{1min}}=&~\FR{\Theta_\text{PN}+\sqrt{\Theta_\text{PN}^2+AC}}{2A},\\
  A=&~(-2+\cos^2I_0)\ep_{10}+\FR{\Theta_\text{PN}}{\sqrt{\ep_{10}}}+4\ep_{10}\bigg(\cos I_0+\FR{\sqrt{\ep_{10}}}{J_{\ga2}/J_{\be1}}\bigg)^2+5,\\
  C=&~20\ep_{10}\bigg(\cos I_0+\FR{\sqrt{\ep_{10}}}{J_{\ga2}/J_{\be1}}\bigg)^2.
\end{align}

 We may want to identify the lifetime of KL-boosted merger $\tau_\text{KL}$ with the lifetime $\tau_\text{iso}$ in (\ref{tiso}) of an imaginary isolated binary with initial ellipticity $\ep_\text{1min}$. But this is not quite right because the inner binary spends only a small portion of its  time around $\ep_\text{1min}$ in each KL cycle, and thus $a_1$ is essentially inert  the rest of the time. The proportion of time around $\ep_\text{1min}$ in each KL cycle can be estimated by asking how long it takes to increase $\ep_1$ from $\ep_\text{1min}$ by $\Delta \ep_1\sim \ep_\text{1min}$. From (\ref{dgdtKL}), we see that the KL time scale is roughly $\dot\ga_1^{-1}\sim J_{\ga1}/K$. To find the time duration that $\ep_{1}$ stays within $(\ep_\text{1min},2\ep_\text{1min})$, we Taylor expand the function $\ep_{1}(t)$ around its minimum $\ep_\text{1min}=\ep_1(0)$ to quadratic order, $\ep_1(t)\simeq \ep_\text{1min}+\ddot\ep_1(0)t^2$, where $\ddot\ep_{1}\sim \ep_\text{1min}\dot\ga_1\sim \ep_\text{1min}K/J_{\ga 1}$ by (\ref{dedtKL}) and (\ref{dgdtKL}). In this way we see that $\ep_1$ stays within $(\ep_\text{1min},2\ep_\text{1min})$ with the time duration $\sim\sqrt{\ep_\text{1min}}J_{\ga 1}/K$ in each KL cycle, or, the inner binary spend only a proportion of $\sqrt{\ep_\text{1min}}$ of the whole KL cycle around $\ep_\text{1min}$. As a result, the merger time of KL-boosted mergers can be estimated as,
\bge
\label{tslow}
  \tau_\text{slow}\simeq\FR{\tau_\text{iso}}{\sqrt{\ep_\text{1min}}}\simeq \FR{5}{256}\FR{c^5a_{10}^4}{G^3m_0m_1m} \ep_{1\text{min}}^{3}.
\ede
We note that this expression also works well for binaries in the isolation limit so long as $e_{10}$ is not very close to 1, for the reasons that $\ep_\text{1min}\simeq 1-e_{10}^2$, and that the ratio between (\ref{tiso}) and (\ref{tslow}) is $\sqrt{1-e_{10}^2}\sim\order{1}$, so the formula for the merger time in (\ref{tslow}) covers both Kozai-boosted mergers and the mergers isolation limit.

\paragraph{Fast mergers.} This is the simplest case as the merger time is simply  the time scale of a KL oscillation, which is the time $\ga_1$ takes to evolve from $0^\circ$ to $90^\circ$ as can be seen from (\ref{dedtKL}). Therefore, we can read the merger time directly from the equation (\ref{dgdtKL}) for $\di\ga_1/\di t$, as,
\begin{align}
\label{tmfast}
  \tau_\text{fast}
  \simeq &~\dot\ga^{-1},\\
  \dot\ga\simeq &~\FR{K}{J_{\ga1}}+\FR{3}{c^2a_{10}\ep_{10}}\Big(\FR{Gm}{a_{10}}\Big)^{3/2}.
\end{align}

\begin{figure}[t]
\centering
\includegraphics[width=0.45\textwidth]{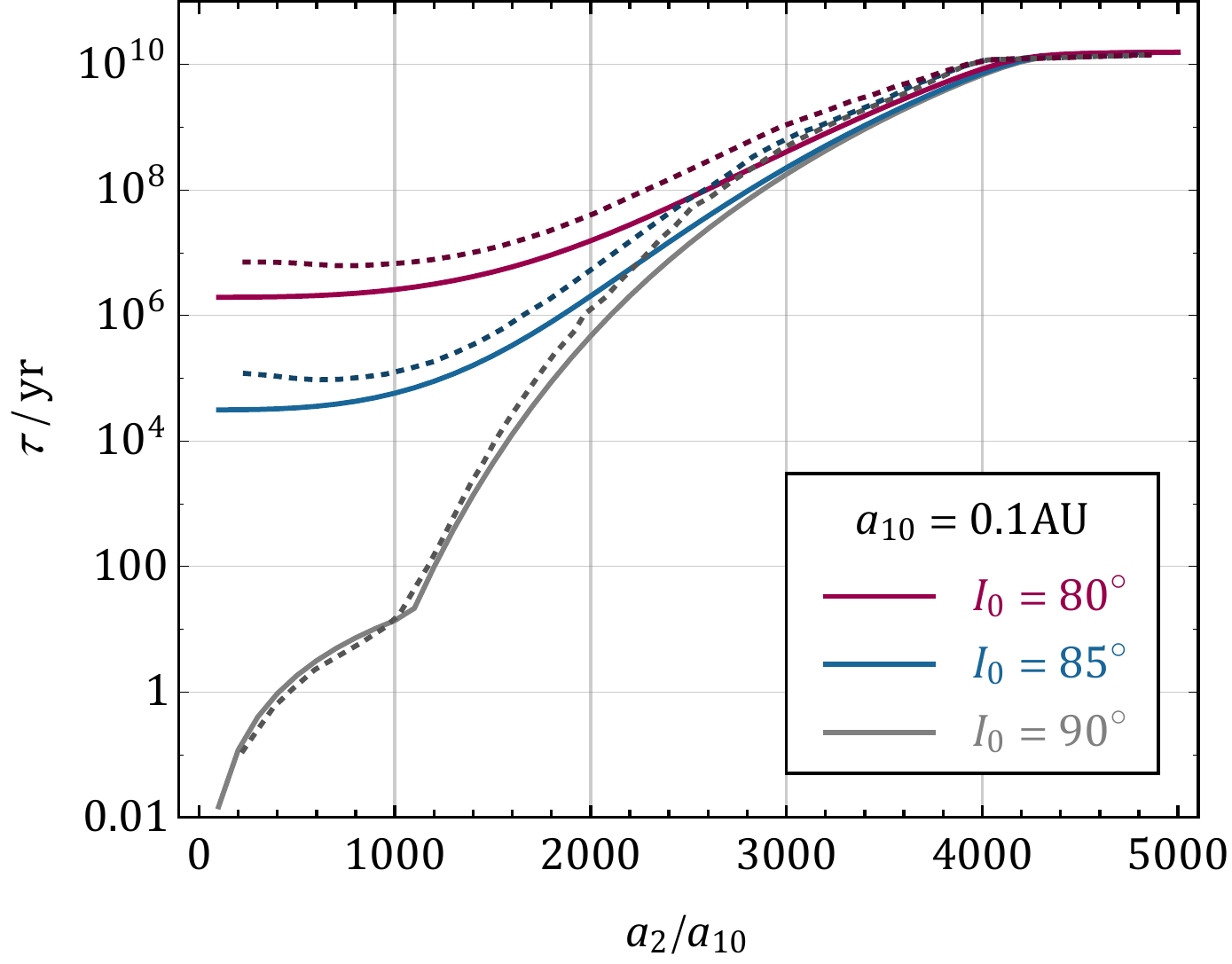}\hspace{3mm}
\includegraphics[width=0.45\textwidth]{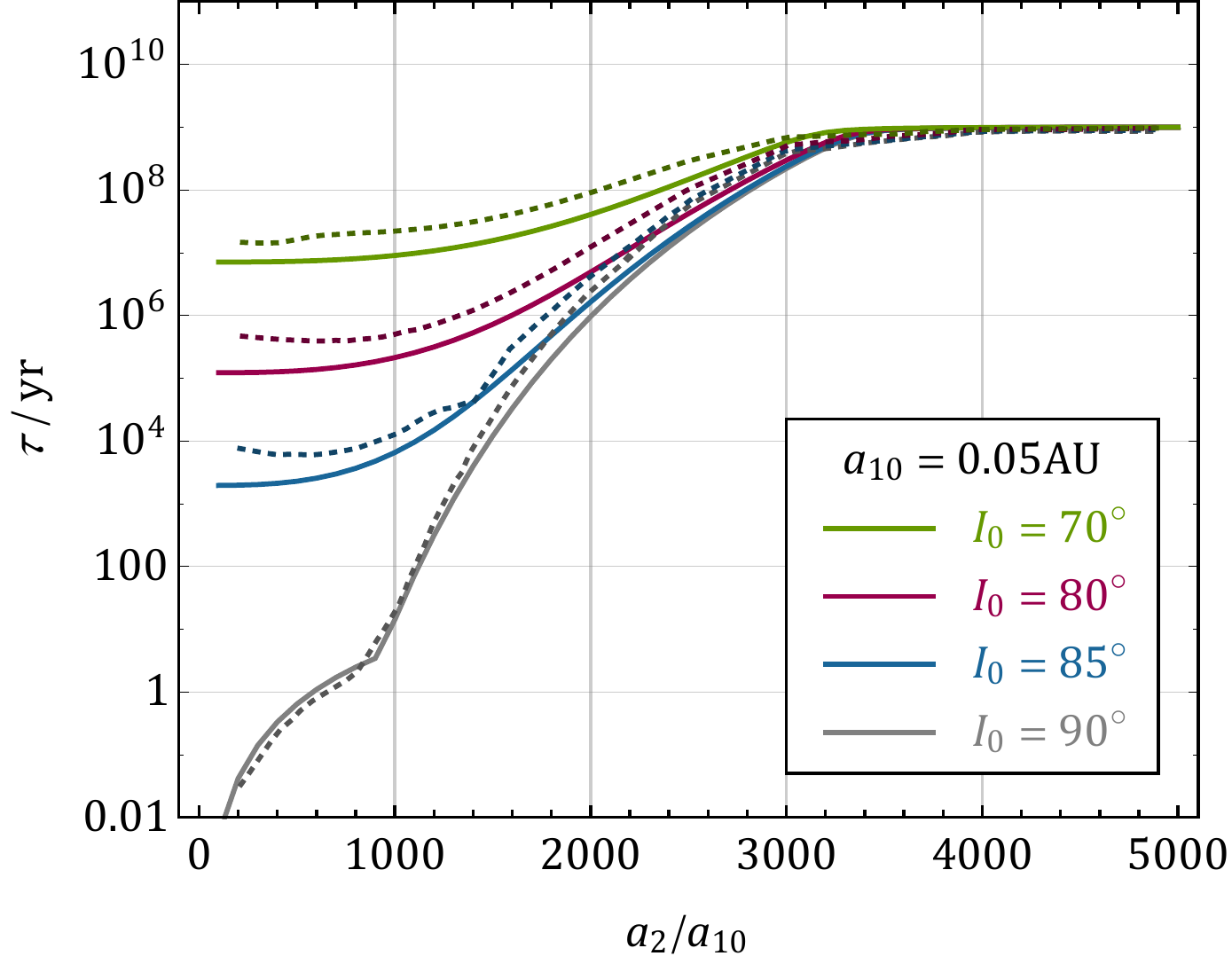}
\caption{The merger time $\tau$ as a function of outer binary separation $a_2$. The parameter choices are, $m_0=m_1=10M_\odot$, $m_2=4\times 10^6M_\odot$, $e_{10}=e_2=0.1$, $\ga_{10}=0$. The inner binary separation $a_{10}$ is set to be 0.1AU and 0.05AU for left and right panels, respectively, while the initial inclination $I_0$ is shown in the plot legend. The solid curves correspond to our analytical estimate (\ref{mergertime}) while the dashed ones are the results of integrating the secular evolution equations (\ref{secularEq}), quoted from \cite{Antonini:2012ad}.}
\label{Fig_tmvsa2}
\end{figure}

Both $\tau_\text{fast}$ and $\tau_\text{slow}$ underestimate the merger time if extrapolated beyond their respective  validity ranges, and therefore the best estimate of the merger time is simply the maximum of the two,
\bge
\label{mergertime}
  \tau=\max\{\tau_\text{fast},\tau_\text{slow}\}.
\ede
In Fig.~\ref{Fig_tmvsa2} we plot the merger time $\tau$ computed from (\ref{mergertime}) as a function of $a_2$ with other parameters fixed. In the same figure we also quote the merger time computed from directly integrating the equations (\ref{secularEq}) as was done in \cite{Antonini:2012ad}. We see the nice agreement between the analytical estimate (\ref{mergertime}) and the direct integration.

\paragraph{KL-boosted, revisited.} Here we provide a complementary estimate of the merger time for KL-boosted binaries. The method used here is technically more involved but we present it because it has a clear physical picture. This estimation uses two assumptions: 1) The reduction of $a_1$ happens only when $e_1$ reaches its maximum in each KL cycle; 2) The maximal $e_1$ doesn't change very much over much of the lifetime of the binary. The second assumption is not very good at late stages so the result derived in this way tends to underestimate the merger time but it is still reasonably good.
 
To proceed, we determine the amount of $a_1$ reduced in each KL cycle, namely the height of each step in the stair-like function $a_1(t)$ in Fig.\;\ref{Fig_Sample2}. Since this happens only in a very narrow range around the moment $t_\star$ where $e_1$ reaches the maximum, namely $e_1(t_\star)=e_{1\text{max}}$, we can approximate $e(t)$ around each $t_\star$ by,
\bge
\label{eExp}
  e_1(t)\simeq e_\text{1max} + \FR{1}{2}\ddot e_1(t_\star)(t-t_\star)^2,
\ede
Here $\ddot e_1(t_\star)$ can be easily obtained from the evolution equations by noting that $\dot e_1(t_\star)=0$,
\begin{align}
  \ddot e_1(t_\star)=&-\FR{10K}{J_{\ga1}}e_{1\text{max}}(1-e_{1\text{max}}^2)\dot \ga(t_\star),\\
  \label{dgdtstar}
  \dot\ga(t_\star)\simeq&~ \FR{K}{J_{\ga1}}+\FR{3}{c^2a_1(1-e_{1\text{max}}^2)}\bigg(\FR{Gm}{a_1}\bigg)^{3/2}.
\end{align}
Then the reduced $a_1$ in each KL cycle can be found by integrating $\dot a(t)$ with $e(t)$ chosen as (\ref{eExp}). The result is,
\begin{align}
  |\Delta a_1|\simeq&~\FR{64G^3m_0m_1 m}{5c^5a_1^3(t_\star)}\Big(1+\FR{73}{24}e_{1\text{max}}^2+\FR{37}{96}e_{1\text{max}}^4\Big)\int_{-\infty}^\infty \FR{\di t}{(1-e_1^2(t))^{7/2}}\n\\
  \simeq &~\FR{1024G^3m_0m_1 m}{75c^5a_1^3(t_\star)(1-e_{1\text{max}}^2)^3}\Big(1+\FR{73}{24}e_{1\text{max}}^2+\FR{37}{96}e_{1\text{max}}^4\Big)\FR{1}{\sqrt{e_{1\text{max}}|\ddot e_1(t_\star)|}}.
\end{align}
Then it remains to estimate the width of the step, which is simply the time scale of KL oscillation, i.e.,
\bge
  (\Delta t)^{-1}\simeq\dot\ga \simeq \FR{K}{J_{\ga1}}+\FR{3}{c^2a_1(1-e_{10}^2)}\bigg(\FR{Gm}{a_1}\bigg)^{3/2},
\ede
which has the same form as (\ref{dgdtstar}) except that it is evaluated with initial $e_{10}$ rather than the maximum $e_{1\text{max}}$.

With both the height and the width of the steps known, we can now write the merger time as,
\begin{align}
\label{tmKozai}
  \tau =\int_0^{a_{10}}\di a_1\,\FR{\Delta t}{|\Delta a_1|},
\end{align}
which essentially agrees with (\ref{tslow}) when $\ep_\text{1min}$ is not too large which we have checked numerically. Indeed, we see that both (\ref{tmKozai}) and (\ref{tslow}) have the same dependence on $a_{10}$ and $e_\text{1max}$ when $e_\text{1max}\sim 1$.

\subsection{Analytical Estimate of Eccentricity}

We are now in a position to figure out the eccentricity distribution of the binary BHs at the time of entering the LIGO window. This amounts to mapping the initial parameter space to the values of eccentricity $e_\text{LIGO}$ at the LIGO threshold, which we take to be 10Hz. The ultimate goal is, given an observed distribution of $e_\text{LIGO}$, to use this map as a way of probing the initial distribution of orbital parameters, and hence the black hole mergers' environments.

The relevant initial orbital parameters include initial values of the inner binary separation $a_{10}$, of the eccentricity $e_{10}$, of the inclination $I_0$, and also of the outer orbital parameters $a_{2}$ and $e_{2}$ which are approximately constant, together with the tertiary mass $m_2$. When we identify the tertiary body to be the SMBH in the galactic center, $m_2$ wil be fixed for a given galaxy. For slow mergers, we can always choose the initial value $\ga_{10}=0$ without loss of generality since we can always wait until $\ga_1$ becomes zero.  
 
To establish the map from the initial parameter space to the eccentricity at the LIGO threshold, it is again helpful to consider fast mergers and slow mergers separately. Once again, by fast mergers we mean those binaries merging within $\order{1}$ KL cycles while slow mergers undergo many KL oscillations. For fast mergers, there is no obvious analytical method to estimate $e_\text{LIGO}$ as the three effects, KL, PN, and GW, are all important in each KL cycle and none can be neglected when estimating $e_\text{LIGO}$. On the other hand, as will be elaborated in the following, we can estimate $e_\text{LIGO}$ for slow mergers quite well using simple analytical methods, since  there are then two distinct stages of  binary evolution,  dominated first by KL oscillations and then by GW circularization.  The second  phase is controlled by the well-known Peters equations (\ref{dadt}) and (\ref{dedt}). We show below that the first state is also analytically tractable to good approximation.  Though slow mergers tend to give rise to small $e_\text{LIGO}$, we show in the appendix that LIGO could in principle be sensitive to such small eccentricities of $\order{0.01}$. By measuring smaller eccentricities, we can probe more of the parameter space. Therefore, we shall take $e_\star=0.01$ as the sensitivity threshold for $e_\text{LIGO}$. More importantly, future GW detectors like LISA could observe such binary BHs at a much earlier stage when their binary separation are several orders of magnitude larger than at LIGO threshold. The eccentricity will then be significantly larger than it is in LIGO since it will have undergone less  GW circularization \cite{Nishizawa:2016eza,Breivik:2016ddj,Michaely:2017fyx}.

We now estimate the eccentricity $e_\text{LIGO}$ for slow mergers. We take an imaginary and isolated binary with the same masses $m_{0,1}$ and initial separation $a_{10}$ as the inner binary in question. Then we set the eccentricity of this imagined isolated binary, denoted by $\wh e_1$, such that it has the same lifetime as the inner binary. Then the eccentricity of the inner binary when entering the LIGO band can be approximated by the eccentricity of the isolated binary at the time of entering the LIGO band. The reason behind this identification is that the lifetime of a slow merger is much longer than the KL time scale, and therefore we can think of the imagined isolated binary as the inner binary with fast KL oscillations averaged away. In the same way, we can think of $\wh e_1$ as an averaged eccentricity of the inner binary over the initial several KL cycles. By the time that the inner binary enters the LIGO band, its KL oscillations have long  ceased due to both PN and GW effects, and therefore, we can identify at this moment the eccentricity of the inner binary by the corresponding value of the imagined and isolated binary.

Putting the above words into equations, we relate $\wh e_1$ with the lifetime $\tau$ of the inner binary by (\ref{tiso}), 
\bge
  \tau=\FR{5}{256}\FR{c^5a_{10}^4}{G^3m_0m_1m}\wh\ep_1^{\;7/2},
\ede
where $\wh\ep_1\equiv 1-\wh e_1^{\,2}$. On the other hand, we know from Sec.~\ref{sec_MergerTime} that the merger time of the inner binary is well approximated by (\ref{tslow}),
\bge 
  \tau=\FR{5}{256}\FR{c^5a_{10}^4}{G^3m_0m_1m} \ep_{1\text{min}}^{3},
\ede
and therefore $\wh\ep_1=\ep_{1\text{min}}^{6/7}$, where $\ep_\text{1min}$ is the minimal value of $\ep_1$ in the first several KL cycles. Then, at a later time when the KL oscillation is totally suppressed, the eccentricity can just be read from the binary separation $a_1$ via (\ref{ge}), namely, 
\bge
\label{eapprox}
  e_1=g^{-1}\bigg[\FR{a_1}{a_{10}}g\bigg(\sqrt{1-\ep_{1\text{min}}^{6/7}}\bigg)\bigg].
\ede
In particular, the eccentricity of the binary at the LIGO threshold, $e_\text{eLIGO}$, is,
\bge
\label{eLIGO}
  e_\text{LIGO}=g^{-1}\bigg[\FR{a_\text{LIGO}}{a_{10}}g\bigg(\sqrt{1-\ep_{1\text{min}}^{6/7}}\bigg)\bigg],
\ede
where $a_\text{LIGO}\simeq 513\text{km}\times(m/M_\odot)^{1/3}$ if the lower end of the LIGO frequency band is taken to be 10Hz. Though this expression looks quite simple, it should be noted that most of the information about the initial condition of the binary is encoded in the value of $\ep_\text{1min}$ in a rather complicated way.

We now can also find the eccentricity distribution $f(e)$ when the binary leaves the tidal sphere of influence, which we define as the region over which KL dominates GWs. In \cite{Randall:2017jop}, this distribution was one of two unknowns required  to find the final eccentricity distribution. Here we explicitly included the PN correction so  we could calculate the final ellipticity distribution directly. Nonetheless, now that we have the expression for the eccentricity at all values of $a_1$ for which the KL oscillation is suppressed, we immediately know that the desired eccentricity can be solved from the expression (\ref{eapprox}) with $a_1$ replaced by its value at the time of exiting the tidal sphere of influence \cite{Randall:2017jop},
\bge
\label{ai}
  a_1=\bigg[a_2^6R_m^5\bigg(\FR{m}{M}\bigg)^2\FR{1}{(1-e_1^2)^6}\bigg]^{1/11},
\ede
where $R_m=2Gm/c^2$ is the Schwarzschild radius associated with binary mass $m$.

\begin{figure}[t]
\centering
\parbox{0.45\textwidth}{\includegraphics[width=0.45\textwidth]{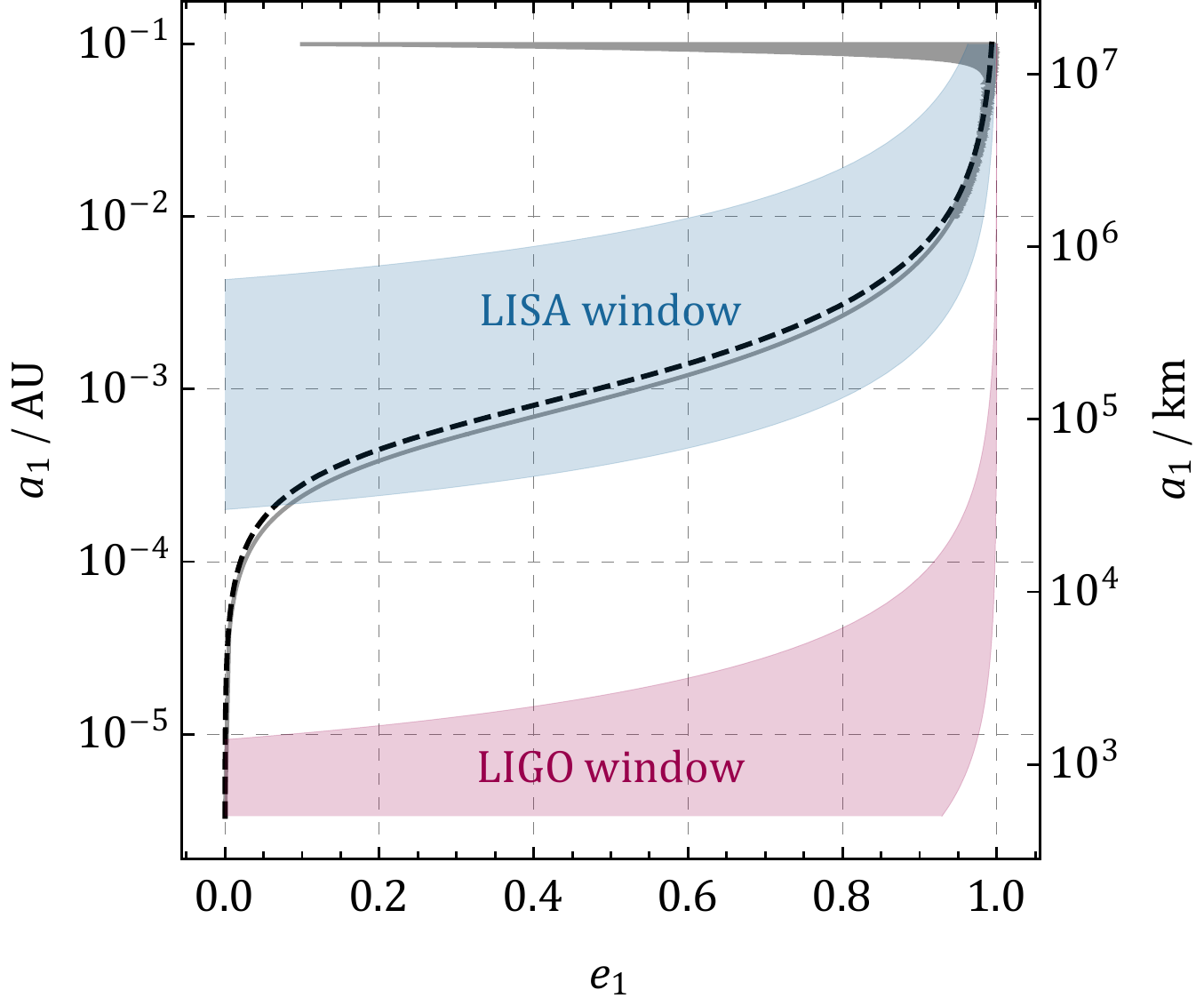}}
\caption{The evolution of BBH in Example 2 in $(e_1,a_1)$ plane. The gray curve is the numerical solution of equations (\ref{secularEq}) and the dashed black curve is the analytical approximation (\ref{eapprox}). The LIGO and LISA windows are taken to be $10\text{Hz}\leq f_\text{GW}\leq 2000$Hz (red) and $0.001\text{Hz}\leq f_\text{GW}\leq 0.1\text{Hz}$ (blue), respectively.}
\label{fig_a1vse1}
\end{figure}

In Fig.\;\ref{fig_a1vse1}, we show the approximate eccentricity (\ref{eapprox}) in the  $(e_1,a_1)$-plane for our second example in Sec.\;\ref{sec_NumExp} (namely the slow merger in Fig.\;\ref{Fig_Sample2}) compared to the numerical solution (the gray curve). It can be seen clearly that the numerical solution undergoes a number of KL oscillations at an early stage while the binary separation stays essentially constant (the upper gray belt). Once the KL oscillation is suppressed, the standard GW circularization takes place and drives the binary to smaller eccentricity. It can be seen that our analytical estimate is a good approximation to $e_1$ for the circularization stage where KL oscillations are suppressed, which is essential to analytically estimate the value of $e_\text{LIGO}$ when the binary enters the LIGO window.

 In the same plot we also show the observational bands of LIGO ($10\text{Hz}\leq f_\text{GW}\leq 2000$Hz) and LISA ($0.001\text{Hz}\leq f_\text{GW}\leq 0.1\text{Hz} $). Both bands are tilted towards large $a_1$ when $e_1$ gets large because the peak frequency $f_\text{peak}$ of emitted GWs moves to larger harmonics as $e_1$ get larger \cite{Wen:2002km}, 
\bge
\label{GWpeak}
f_\text{peak}(a_1,e_1)\simeq \FR{\sqrt{Gm}}{\pi[a_1(1-e_1^2)]^{3/2}}(1+e_1)^{1.1954}.
\ede 
Since LISA is able to probe stellar BBHs with much greater separation, it can in principle capture more information stored in eccentricity before the circularization washes it away. In the future, it is possible that a compact binary enters both LISA and LIGO windows at different stages of inspiral, and a joint analysis with both ground and space GW detectors can be very powerful in measuring the properties of inspiral binaries, and a LIGO event without LISA counterpart can also provide us useful information about the inspiral history \cite{Nishizawa:2016eza,Breivik:2016ddj,Michaely:2017fyx}.

\begin{figure}[t]
\centering
\parbox{0.35\textwidth}{\includegraphics[width=0.35\textwidth]{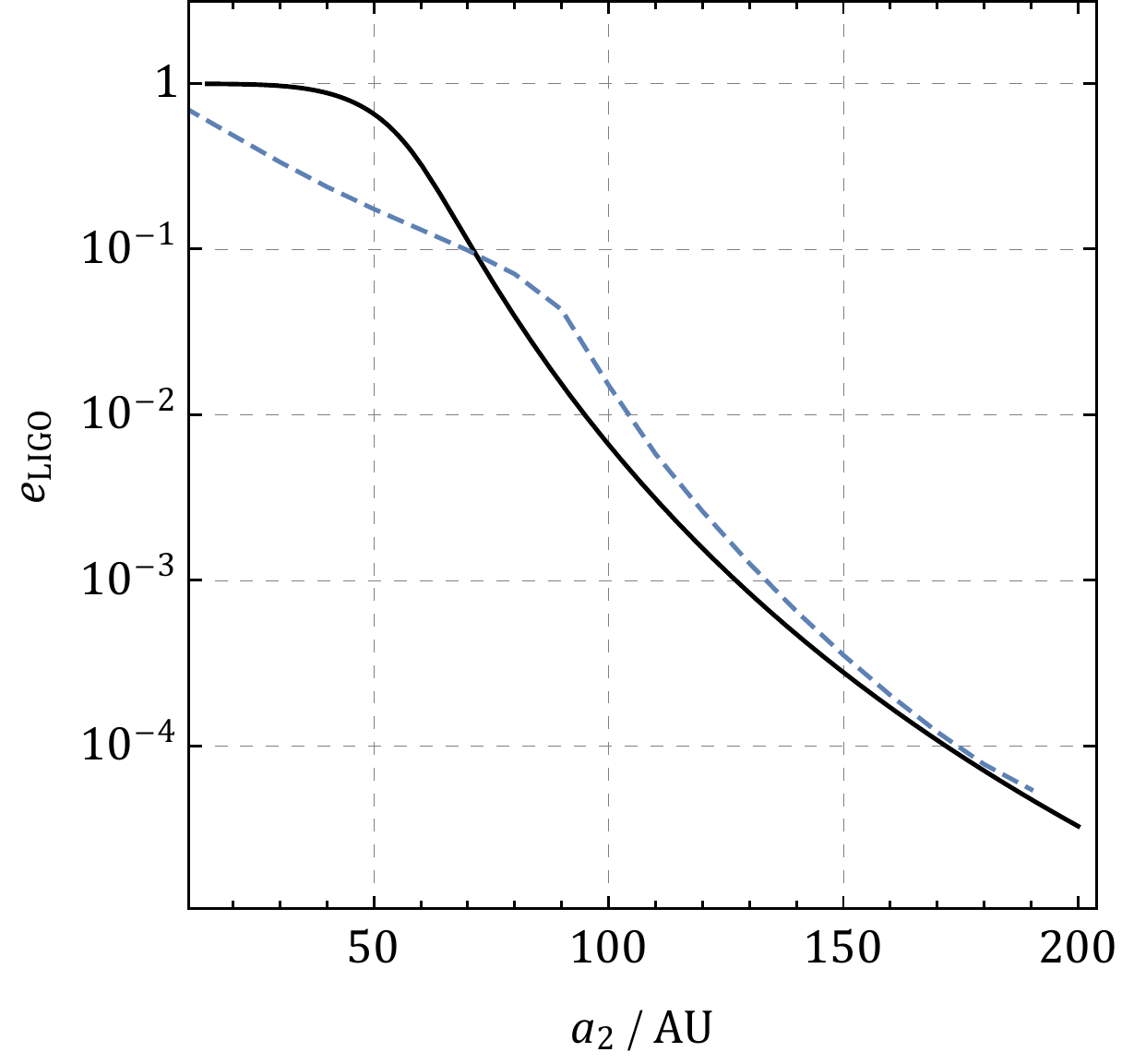}}
\caption{The eccentricity of an inner binary when entering the LIGO band as a function of the distance $a_2$ to the central SMBH. The solid black curve and dashed blue curve correspond to the analytical estimate (\ref{eapprox}) and the numerical solution, respectively. All initial parameters except $a_2$ are taken to be the same with the three examples in Sec.\;\ref{sec_NumExp}.}
\label{fig_eLIGOvsa2}
\end{figure}

In Fig.\;\ref{fig_eLIGOvsa2} we compare the numerical solution and analytical estimate (\ref{eLIGO}) for the eccentricity $e_\text{LIGO}$ of the BBH at LIGO threshold. We vary $a_2$ to show different strength KL oscillations  while all other parameters are taken to be the same as in the three examples in Sec.\;\ref{sec_NumExp}. As expected, the estimate (\ref{eLIGO}) works quite well for small eccentricity due to a large number of KL oscillations and a clear distinction between KL-domination and GW-domination. At large eccentricity the estimate (\ref{eLIGO}) does not agree with the numerical solution as well, but still serves as a good indication in the order-of-magnitude sense. In general, large eccentricities correspond to fast mergers, and such events are expected to be rarer than ones with smaller eccentricity in the galactic center since they happen only in the small-volume inner region that is only slowly replenished.

\section{Eccentricity Distribution}
\label{sec_ED}

Now that we can calculate the final eccentricity distribution given a set of initial parameters, we consider in this section the expected distribution in a NC with a central SMBH. The ultimate physical goal would be to use measured distributions to determine if this is indeed the origin of the black hole merger, and the parameters of the initial binaries in the case where they merge sufficiently close to the SMBH for it to influence their orbit, ultimately telling about the density distribution in the central region.  This is of course hampered by the large number of parameters and the associated degeneracies, as well as the limited statistics that will be available. However, we will see that the result depends primarily on only a few of the initial parameters, namely $m_2$, $e_2$, $a_2$, and $I$ as well as other parameters which are measured essentially directly,  $m_0$ and $m_1$. Although statistics are of course currently very limited, the hope is that over time we will have enough measured binary mergers to get a more detailed understanding of the density profile $\rho (a_2)$ describing the distribution of binaries as a function of distance from a central black hole. On top of this, we have focused only on the LIGO potential so far. In conjunction with future measurements such as the ones that should be possible from eLISA, we will have more and better measurements of the mergers.

\subsection{Parameter Dependence}
\label{sec_par_dep}

As we showed in Sec.\;\ref{sec_merger} the final eccentricity $e_\text{LIGO}$ depends on a number of parameters, including the two masses of the binary $(m_0,m_1)$, the mass of the central SMBH $m_2$, the semi-major axis and eccentricity of the outer orbit $(a_2,e_2)$, the initial semi-major axis and the eccentricity of the inner binary $(a_{10},e_{10})$, and  the inclination $I_0$ of the inner orbit relative to the outer orbit.

Among these parameters, only the binary masses $(m_0,m_1)$ are directly measurable by LIGO, and thus we do not know most of them for each event. However, these parameters can leave their impact on the final eccentricity $e_\text{LIGO}$, and the hope is that we will obtain useful information about these parameters with more statistics of observations. To this end, it is important to understand how the parameters affect the distribution of $e_\text{LIGO}$ and how well we know about their initial distribution. We comment on these dependencies now.

In general, most parameters enter the formula (\ref{eLIGO}) for $e_\text{LIGO}$ through the value of $e_\text{1max}$, the maximal eccentricity in the first several KL cycles. Therefore, their importance to the final $e_\text{LIGO}$ basically depends on how much they affect the value of $e_\text{1max}$.

The final distribution of $e_\text{LIGO}$ does not depend strongly on $e_{10}$, and has very mild dependence on $a_{10}$. For $e_{10}$, this is because the information about initial eccentricity is largely washed out by many KL cycles, unless $e_{10}$ is extremely large. However, binaries with  extremely large $e_{10}$ are rare, because we expect that the distribution of $e_{10}$ in very dense environment to be flat in $e_{10}^2$, which follows from a equipartition argument firstly given by James Jeans \cite{Jeans1919}. 

On the other hand, we know very little about the $a_{10}$ distribution in a NC. But we do not need detailed knowledge about it either because binaries with very large $a_{10}$ will evaporate anyway, and very small $a_{10}$ is hard to form. In typical NCs, this means that we can assume that $a_{10}$ follows some distribution ranging from $\order{0.01\text{AU}}$ to $\order{10\text{AU}}$, and we find that the distribution of $a_{10}$ in this range has only a tiny impact on the final answer for $e_\text{LIGO}$. Therefore we will assume a flat distribution for $a_{10}$ in the following.

The most influential parameter affecting the final eccentricity distribution is the inclination $I_0$ because it largely determines $e_\text{1max}$ in the absence of the PN correction. We will assume that the orientation of the inner binary follows a random uniform distribution in all possible directions --- that is we assume a flat distribution in $\cos I_0$. This is also the assumption adopted in \cite{Antonini:2012ad}. Eventually, binaries with low inclination will evaporate since they receive little KL oscillations and thus do not merge fast enough. We will take account of the evaporation constraint explicitly shortly.  

After including the PN correction, the outer orbital parameters $(a_2,e_2)$ become important because they affect the rate of KL oscillations, and this rate competes with the PN correction in determining $e_\text{1max}$. 
The distribution of $e_2$ is important, because a relatively large $e_{2}$ can reduce the distance of periapsis for fixed $a_2$ and the binary may receive larger perturbations there. In our examples we follow \cite{Antonini:2012ad} and take the distribution of $e_2$ to be thermal, i.e. $\propto \di e_2^2$. although ultimately this is an assumption that should be observationally checked. 

We know little about the $a_2$ distribution in a NC, and finding a method for observationally determining this is potentially one of the most important goals of the approach we describe in this paper since at this point it is extremely difficult to probe this distribution close to the galactic center in other ways.  Examples of possible distributions as described in the literate are:  For less relaxed systems with a core profile for the stellar distribution, one possibility is that the distribution of $a_2$ follows the background. As in \cite{Antonini:2012ad}, we take the profile to be $a_2^{2-\be}\di a_2$ with $\be=0.5$ for the core model. For the fully relaxed case with a Bahcall-Wolf cusp, it is expected that the BH binaries would be more cuspy due to mass segregation if the BH masses are much greater than the background stellar mass when the binary BH is not the dominant component of the NC  \cite{Alexander:2008tq}. In this case we shall take the mass-segregated distribution with $\be=2$. But for lighter binaries with $m\sim M_\odot$, there is no segregation at work and we shall take $\be=7/4$ to be in accordance with the Bahcall-Wolf cusp. Throughout we restrict $a_2$ within the inner 0.1pc from the central SMBH. 

This leaves the most important remaining parameter as $I$. As discussed, we take a flat distribution initially. However, mergers compete with evaporation in determining the fate of black hole binaries, which favors large inclinations for successful mergers. So the final consideration we apply is comparing merger and evaporation rates, the latter of which is already discussed in the literature and we review in the next section.

\subsection{Evaporation Rate} 

In dense environments such as NCs, BH binaries can be destroyed by background stellar mass objects through binary-single interactions and thus ``evaporate'' rather than merge \cite{Leigh:2017wff}. Therefore to be observed by LIGO or other detectors it is important  that the merger time is shorter than the evaporation time scale. This introduces a constraint on the parameter space of the initial distribution of binaries. This constraint was taken into account numerically in previous studies \cite{Wen:2002km,Antonini:2012ad}. To impose the constraint analytically, we use  the cross section of the process $\si(\text{binary}+\text{single}\to \text{three~singles})$ in conjunction with the analytical merger calculation we did above. 

Intuitively, the evaporation cross section can at most be the geometrical cross section, $\si\sim\pi a_1^2$, according to which binaries with large separation evaporate more readily. The cross section also depends crucially on the velocity of the scattered object $v_\star$ relative to the binary mass center. In fact, the geometrical cross section $\pi a_1$ can be reached only when $v_\star$ is mildly greater than the rotational velocity of the binary $v_b\sim\sqrt{Gm/a_1}$, because the time duration of the close encounter is too short to destroy the binary if $v_\star\gg v_b$, and there is not enough energy to evaporate the binary if $v_\star\ll v_b$.

A binary is conventionally said to be ``hard'' if $v_b > v_\star$ and ``soft'' if $v_b < v_\star$. In galactic nuclei with a SMBH, the typical velocity of field stars is $v_\star\sim \sqrt{Gm_2/a_2}$ with $m_2$ and $a_2$ the mass of and the distance to the central SMBH, respectively. Therefore a binary is hard if $a_1\lesssim (m/m_2)a_2$. For typical masses $m\sim 10M_\odot$ and $m_2\sim 10^6M_\odot$, a binary in the central region $a_2 < 10^4$AU of a NC will in general be soft if $a_1>0.1$AU. Hard binaries do not evaporate. For very soft binaries, an analytical expression for the cross section of evaporation is known \cite{Hut1983}, 
\bge
  \si_\text{evap}=\FR{40\pi G}{3}\FR{m_\star^2}{m}\FR{a}{v_\star^2},
\ede
where $m_\star$ is the mass of the field star and $v_\star$ is its velocity relative to the binary. From this expression we can find the evaporation time scale $\tau_\text{evap}$ of the binary,
\bge
\label{tevap-1}
  \tau_\text{evap}^{-1}=n_\star \la \si_\text{evap}v_\star \ra=\FR{40\pi Ga}{3}\FR{\rh_\star m_\star}{m}\big\la v_\star^{-1} \big\ra,
\ede
where $n_\star$ and $\rh_\star$ and number density and mass density of the background stars, respectively, and the average $\la\cdots\ra$ is over the velocity distributions with a cutoff at $v_b$. We assume that the velocity of field stars follows the Maxwell distribution, $f(v)=e^{-v^2/(2\bar v^2)}v^2\di v$. Then the variance of the velocity distribution $\bar v$ can be determined in terms of densities by solving the Jeans equation.   At distances $a_2$ close to the SMBH its mass dominates, so the velocity dispersion is approximately 
\bge
  \bar v^2 \sim \FR{GM_\text{SMBH}}{a_2}. 
\ede
Additional velocity contributions \cite{1608.02944} due to stellar mass are subdominant for the parameters of interest here. 
Then, the evaporation time scale is given by,
\bge
\label{tevap}
  \tau_\text{evap}=\FR{3m\bar v }{40\sqrt{2\pi}G\rh_\star m_\star a_{10}}.
\ede
We now impose the constraint $\tau_\text{evap}>\tau$ where $\tau$ is given in (\ref{tslow}). In the situation considered in \cite{Randall:2017jop} where the initial binary separation is fixed and the inclination is assumed to be large enough, this evaporation constraint just reduces to a cutoff on $a_2$, which was one of two unknowns in \cite{Randall:2017jop} but which we can now obtain analytically. 
 
The background density profile $\rh_\star$ makes an impact on the final result through the evaporation constraint, which means that the distribution of $e_\text{LIGO}$ can be sensitive to the background profile. In the central region of a NC, we assume that the density has a power-law profile,
\bge
\label{rhostar}
  \rh_\star(a_2)=\rh_0\bigg(\FR{a_2}{a_{20}}\bigg)^{-\al},
\ede
where $\rh_0$ is the density at a benchmark distance $a_{20}$.
We use models for the density profile in the literature, though our analysis would apply to any proposed profile.  As mentioned earlier, we take the Bahcall-Wolf cusp profile with $\al=7/4$ for fully relaxed galaxies and a core model with $\al=1/2$ for less relaxed galaxies.

{\subsection{Sample Results}

\begin{figure}[t]
\centering
\parbox{0.3\textwidth}{\includegraphics[height=0.28\textwidth]{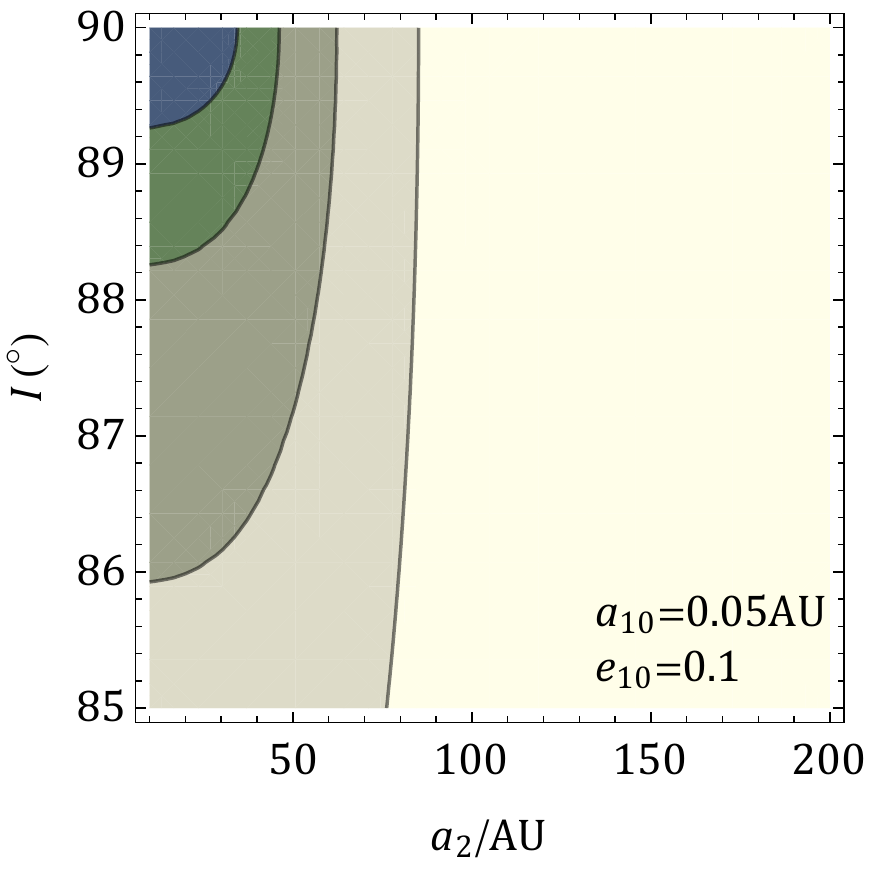}}\hspace{2mm}
\parbox{0.3\textwidth}{\includegraphics[height=0.28\textwidth]{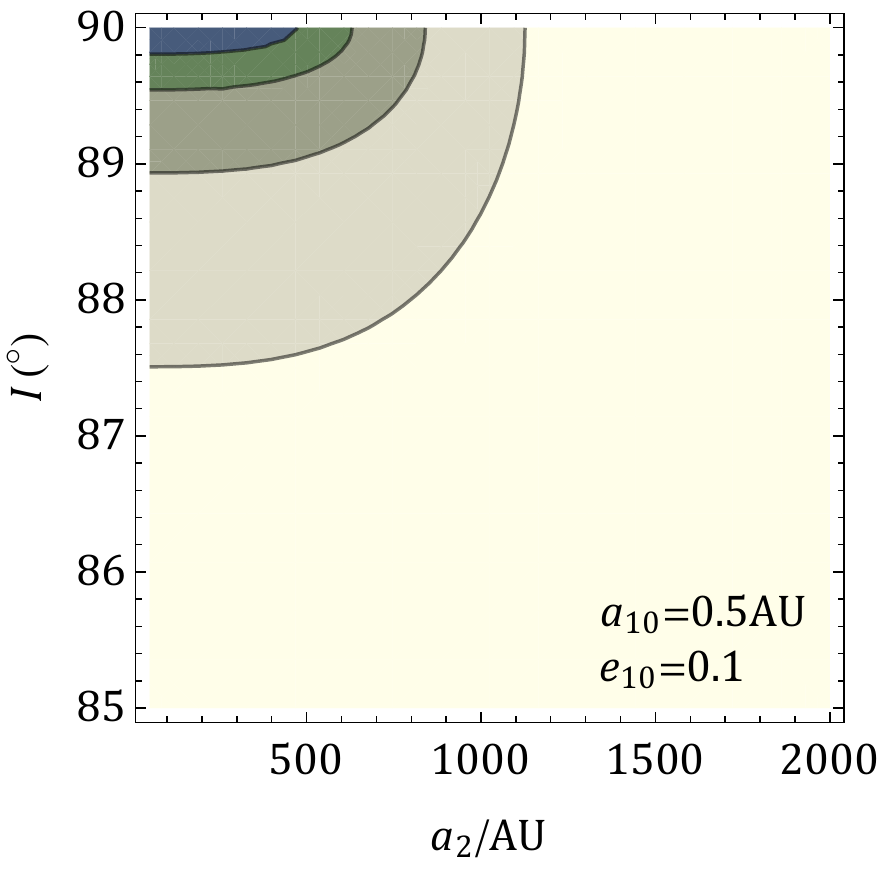}}\hspace{2mm}
\parbox{0.3\textwidth}{\includegraphics[height=0.28\textwidth]{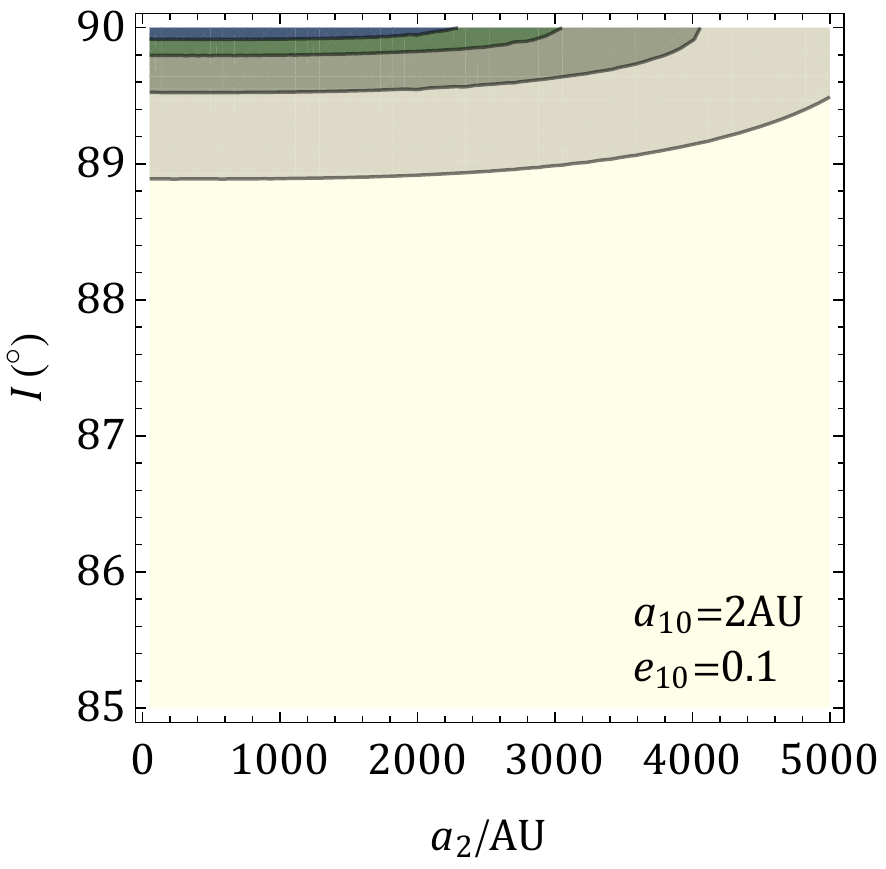}}\\
\parbox{0.3\textwidth}{\includegraphics[height=0.28\textwidth]{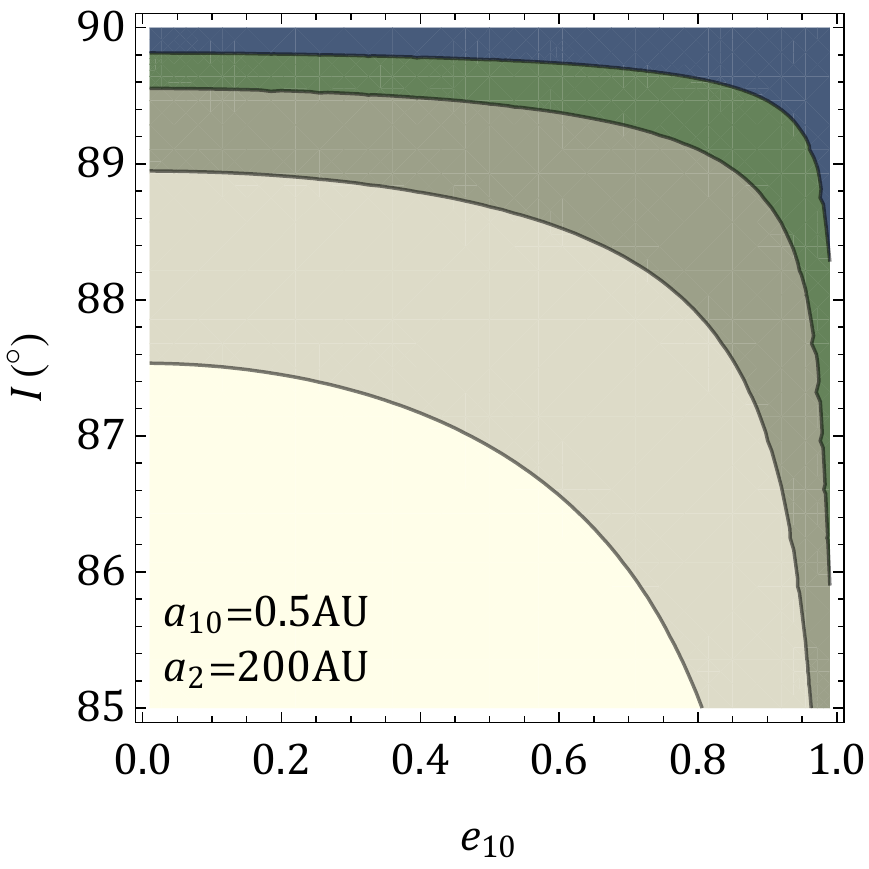}}\hspace{2mm}
\parbox{0.3\textwidth}{\includegraphics[height=0.28\textwidth]{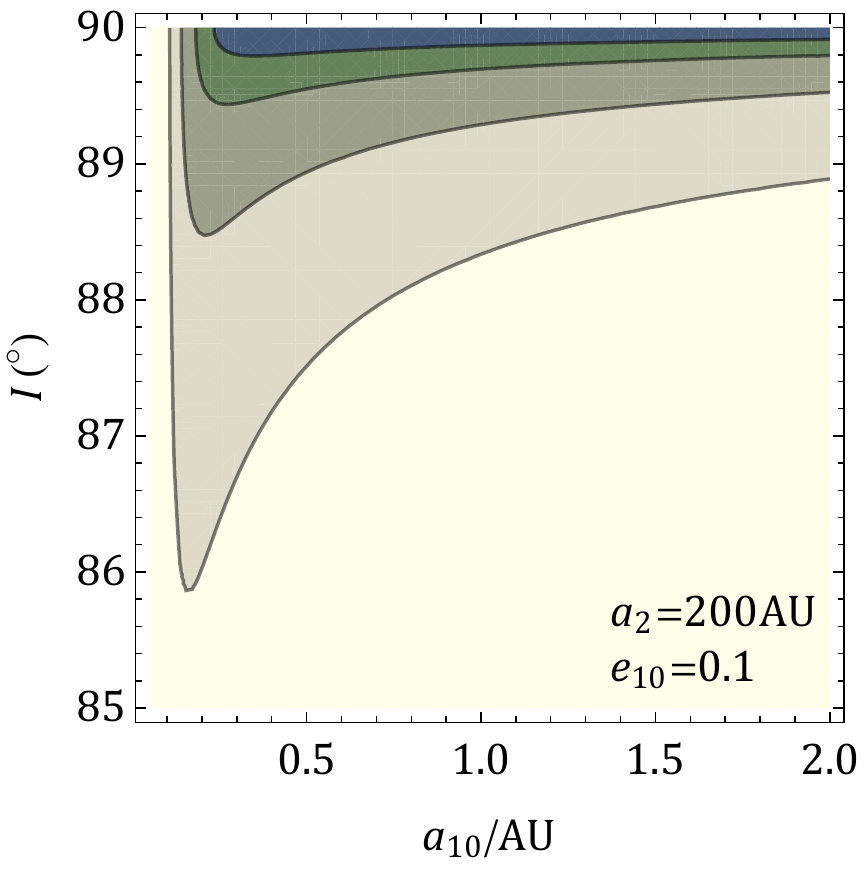}}\\
$e_\text{LIGO}:$ \parbox{0.35\textwidth}{\includegraphics[width=0.35\textwidth]{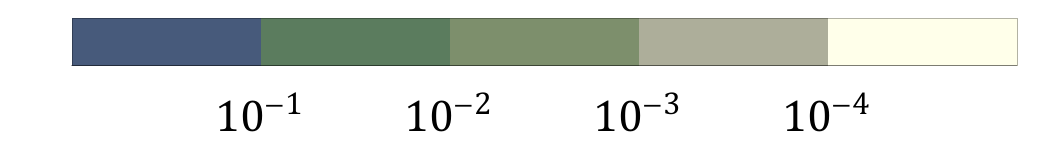}}
\caption{The eccentricity of an inner binary when entering the LIGO band with various initial conditions typical in galactic nuclei. In all plots $m_0=m_1=10M_\odot$, $m_2=4\times 10^6M_\odot$, and $e_2=0.1$. The two panels in the second row show the weak dependence of $e_\text{LIGO}$ on $e_{10}$ and $a_{10}$.}
\label{fig_eDist_NC}
\end{figure}

To conclude this subsection, let us consider some limits of the expression (\ref{eLIGO}) to gain a bit of insight into the eccentricity $e_\text{LIGO}$. Here we consider instances where the final eccentricity is small but not unobservable, i.e. $e_\text{LIGO}\sim 0.01$ to 0.1. We assume that $a_{10}\gg a_\text{LIGO}$ so that $\ep_\text{1min}$ is small. With these assumptions, we can use the $e\simeq 1$ limit of $g(e)$ in (\ref{ge}) for the inner layer of (\ref{eLIGO}) and use the $e\ll 1$ limit for $g^{-1}$. As a result, we have,
\bge
\label{eLIGOapp}
  e_\text{LIGO}\simeq 1.22\bigg(\FR{a_\text{LIGO}}{a_{10}}\bigg)^{19/12}\ep_\text{1min}^{-19/14}.
\ede
Let us now consider the limit of a highly inclined inner binary with $I\simeq 90^\circ$ so that $\ep_\text{1min}$ in (\ref{eminsol}) is dominated by $\Theta_\text{PN}$. In this case we have $\Theta_\text{PN}^2\gg AC$ and $A\simeq 3$ in (\ref{eminsol}), and thus $\ep_\text{1min}\simeq \Theta_\text{PN}^2/9$. We also use $a_\text{LIGO}\simeq 513\text{km}\times(m/M_\odot)^{1/3}$, which is true for small $e_\text{LIGO}$. Putting these two expressions into (\ref{eLIGOapp}), we get,
\begin{align}
\label{eLIGO_largeI}
  e_\text{LIGO}\simeq 0.01\bigg(\FR{20M_\odot}{m}\bigg)^{1235/252}\bigg(\FR{m_2}{4\times 10^6M_\odot}\bigg)^{19/7}\bigg(\FR{a_{10}}{0.1\text{AU}}\bigg)^{779/84}\bigg(\FR{100\text{AU}}{b_2}\bigg)^{57/7},
\end{align}
where $b_2=a_2\sqrt{1-e_2^2}$ appears from $W_\text{PN}$ in (\ref{WPN}). We learn from this expression that the binary gains more eccentricity with smaller binary mass, larger central SMBH, and shorter distance to the central SMBH. The eccentricity increases with $a_{10}$ because tidal force is stronger for larger binary separation.  

As another interesting limit, we take the inner binary to be very close to the tertiary body but with mildly large inclination, which means that the dominant contribution to $\ep_\text{1min}$ in (\ref{eminsol}) is the not-so-large inclination angle. From (\ref{WPN}) we see that $\Theta_\text{PN}\propto a_2^3$ and thus the PN effect is suppressed for very small $a_2$, so we have $AC\gg\Theta_\text{PN}$ and we can neglect $\Theta_\text{PN}$ in (\ref{eminsol}) and use the approximation $A\simeq 3+5\cos^2I_0$ and $C\simeq 20\cos^2I_0$, and thus $\ep_\text{1min}\simeq 5\cos^2 I_0/(3+5\cos^2 I_0)$. Using (\ref{eLIGOapp}) again, we get,
\bge
\label{eLIGO_smalla2}
 e_\text{LIGO}\simeq 0.01\bigg(\FR{m}{20M_\odot}\bigg)^{19/36}\bigg(\FR{0.1\text{AU}}{a_{10}}\bigg)^{19/12}\bigg(\FR{\cos 88.8^\circ}{\cos I_0}\bigg)^{19/7}.
\ede
The $m$ and $a_{10}$ dependence in (\ref{eLIGO_smalla2}) is from the $a_\text{LIGO}$ dependence in (\ref{eLIGOapp}), and is in opposite trend to (\ref{eLIGO_largeI}). As we will see later, the final distribution of $e_\text{LIGO}$ will be anti-correlated with $m$ because (\ref{eLIGO_largeI}) has stronger $m$-dependence than (\ref{eLIGO_smalla2}) and also because more elliptical events are from the high-inclination limit (\ref{eLIGO_smalla2}) than from the small-$a_2$ limit (\ref{eLIGO_smalla2}). Clearly the $m$-dependence depends on where the events arise and merits further study as it can ultimately be very interesting in studying the black hole merger environments. 

We show in Fig.~\ref{fig_eDist_NC} the eccentricity $e_\text{LIGO}$ at the LIGO threshold in various sections of the initial parameter space, using the analytical estimate (\ref{eLIGO}). From the three panels in the first row, we can see that larger-$e_\text{LIGO}$ regions (with darker shade) move from left side to the top boundary as $a_{10}$ increases from a smaller value to a larger value. This means that harder binaries get large eccentricities when they are closer to the central SMBH compared with softer ones, and that softer binaries could reach large eccentricities when they are far away but have very large inclinations. The lower-left sides of these contours correspond to the limit described by (\ref{eLIGO_smalla2}), while the upper-right sides correspond to the limit of (\ref{eLIGO_largeI}).

In the second row, the left panel shows that $e_\text{LIGO}$ is not sensitive to the initial eccentricity $e_{10}$ unless $e_{10}$ is very large, and the right panel shows that $e_\text{LIGO}$ is not sensitive to initial separation $a_{10}$ either, unless $a_{10}$ is extremely small. Finally, the right panel shows that highly inclined binaries can be very eccentric.

\subsection{Examples of Eccentricity Distribution}

We are now ready to map binaries prepared with a given distribution of initial parameters to their distribution of eccentricity $f_\text{LIGO}(e)$ at the LIGO threshold. For fixed binary masses $(m_0,m_1)$ and mass of the central SMBH $m_2$, the initial parameters include the binary separation $a_{10}$ and eccentricity $e_{10}$, the semi-major axis $a_{2}$ and eccentricity $e_2$ of the outer orbit, and the mutual inclination angle $I_0$. Given the analytical expression (\ref{eLIGO}), we are in principle free to explore any distribution of initial parameters. 

In our examples here, we fix $m_2=4\times 10^6M_\odot$, and consider two cases for the inner binary masses: one with $m_0=m_1=10M_\odot$ and the other with $m_0=m_1=M_\odot$. For the inner orbital parameters, we take $e_{10}=0$ since the final answer is not sensitive to this value, and we take a uniform distribution $\propto \di a_{10}$ ranging from $0.1$AU to $5$AU. For the outer orbital parameters, we will take a thermal distribution for the eccentricity $\propto \di e_2^2$, and take the semi-major axis distribution to be $a_2^{2-\be}\di a_2$ with $\be=2$ for the cusp model and $\be=0.5$ for the core model. 
 
We can now  integrate over the parameter space spanned by $(a_{10},a_2,e_2,I_0)$ for fixed masses $(m_0, m_1, m_2)$, where we assume the mass parameters $m_0$ and $m_1$ can be independently determined and $m_2$ is a fixed parameter in a given galaxy. To find the final eccentricity distribution at the LIGO threshold, we note that our analytical equation (\ref{eLIGO}) foliates this parameter space into equal-$e_\text{LIGO}$ slices, and the probability of finding a binary with eccentricity between $e$ and $e+\Delta e$ is proportional to the integral over the slice between the $e$-surface and the $(e+\Delta e)$-surface, weighted by the distribution of initial parameters as elaborated above.

 On top of this, we also impose the constraint of non-evaporation, $\tau<\tau_\text{evap}$ as discussed before, and the constraint of binary stability \cite{Randall:2017jop}, 
\bge
  a_2(1-e_2)>\bigg(\FR{4m_2}{m}\bigg)^{1/3}a_{10},
\ede
which says that the binary should not be too close to the central SMBH when it reaches the periapsis or  the tidal force would overwhelm the self-gravity of the binary and disintegrate it.  Comparing with (\ref{weakcondition}), we see this condition ensures the weak coupling between the inner and outer orbits, and thus the stability of the hierarchical triple system.

In summary, the probability $P(e,\Delta e)$ that a BH binary near a SMBH enters the LIGO window with eccentricity between $e$ and $e+\Delta e$ is given by,
\begin{align}
\label{Pe}
  P(e,\Delta e)=\FR{\int_{e_\text{LIGO}\in (e, e+\Delta e)} \di x\,f_\text{ini}(x) C(x) }{\int \di x\,f_\text{ini}(x) C(x)},
\end{align}
where $x=(a_{10},e_{10},a_2,e_2,I_0)$ denotes compactly the initial parameters, $f_\text{ini}(x)$ represents the initial distributions of $x$, while $C(x)$ contains the two constraints that can be expressed in terms of Heaviside step function $\theta(x)$,
\bge
  C(x)\equiv\theta(\tau_\text{evap}-\tau)\theta\big[a_2(1-e_2)-(4m/m_2)^{1/3}a_{10}\big].
\ede

\begin{figure}[t]
\centering
\includegraphics[width=0.45\textwidth]{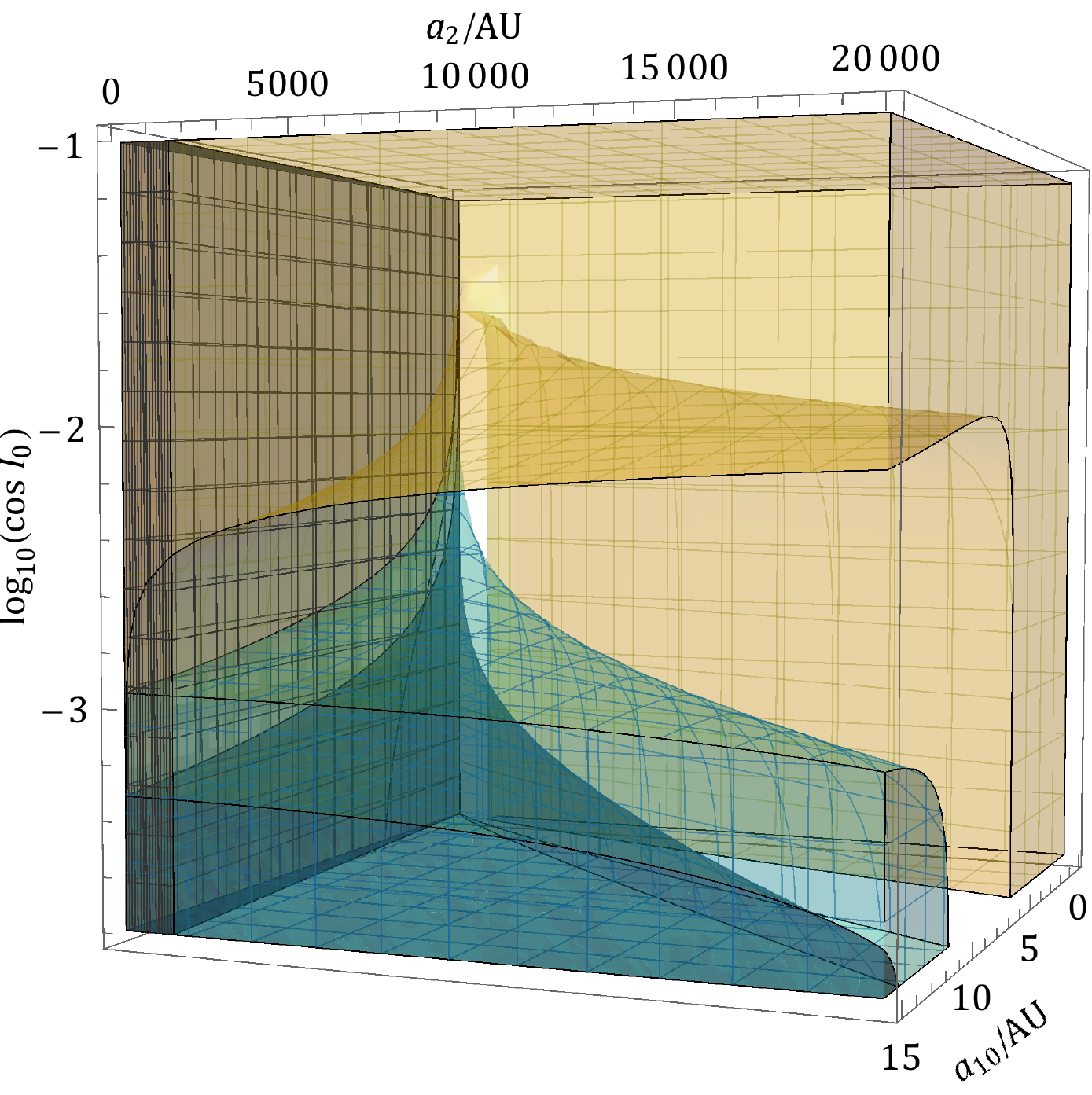}
\caption{A projection of the initial parameter space $(a_{10},a_2,\cos I_0)$ with $e_{10}=e_2=0$. The teal blue regions correspond to parameters leading to merger with eccentricities larger than 0.1 (darker and lower shaded region) and 0.01 (lighter and upper shaded region). The yellow regions lead to evaporation while the gray regions correspond to tidal disruption.}
\label{fig_eDist3d}
\end{figure}
In Fig.\;\ref{fig_eDist3d} we take $m_0=m_1=10M_\odot$, $m_2=4\times 10^6M_\odot$ and show two layers of equal-$e_\text{LIGO}$ slices (in teal blue) for $e_\text{LIGO}=0.01$ (upper and light) and $e_\text{LIGO}=0.1$ (lower and dark) in the parameter space of $(a_{10},a_2,\cos I_0)$ with $e_{20}=0$. Also shown are regions excluded by the evaporation constraint (yellow) and tidal disruption (gray). The evaporation constraint is computed from the cusp model (\ref{rhostar}) with $m_\star=M_\odot$, $\al=7/4$, $\rh_0=10^6M_\odot/\text{pc}^3$ and $a_{20}=0.1$pc.

We perform the integration (\ref{Pe}) for several sets of initial distributions and show the resultant probability $P(e)$ in Fig.\;\ref{fig_eHisto}. For the cusp model, we take the same background profile as the one we take for Fig.\;\ref{fig_eDist3d}, while the core profile corresponds to replacing $\al=7/4$ by $\al=1/2$. It is clear from the figure that the cusp profile tends to produce a more elliptic distribution than a core model when other parameters are fixed. More interestingly, lighter binaries tend to gain more eccentricity in NC than heavier binaries, which means there is an anti-correlation between the binary mass and eccentricity in this formation channel. 

This is different from the binaries in GCs where the mass has little impact on eccentricity distribution \cite{Wen:2002km}, and is in contrast to what is claimed in \cite{Gondan:2017wzd} who considered an alternative eccentric BBH formation channel with direct two-body encounter and found that the binary mass and the eccentricity is positively correlated. Though we have yet to analyze such situations, such parameter-dependence might ultimately be used to distinguish different formation scenarios. 

In general, it is clear that most binaries in NCs will  have small eccentricities. A careful measurement of eccentricity distribution in this range could be very important in revealing the formation of binaries. It is likely that LIGO can  search  only for the tail of this distribution in the not-too-small $e$ region ($e\geq 0.01$, see App.\;\ref{app_small_e}), which contributes $5\%$ ($24\%$) of all mergers in NCs in the cusp model with $m=20M_\odot$ ($m=2M_\odot$). Further, a joint observation with future GW detectors such as LISA would be a lot more powerful in measuring the distribution in the region of tiny (LIGO) eccentricities, and it is even possible to reveal the peak in the distributions if this formation channel contributes significantly to total merger events.

\begin{figure}[t]
\centering
\includegraphics[height=0.36\textwidth]{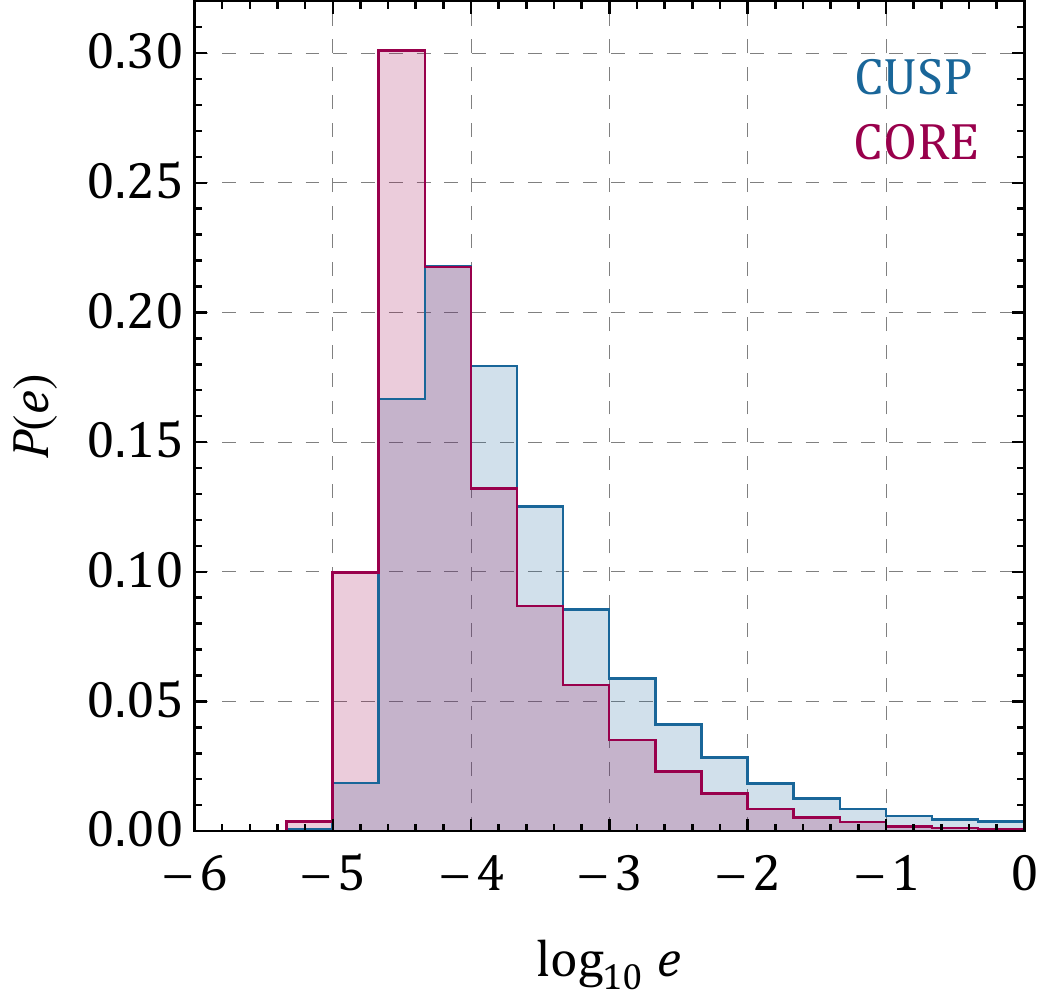}\hspace{5mm}
\includegraphics[height=0.36\textwidth]{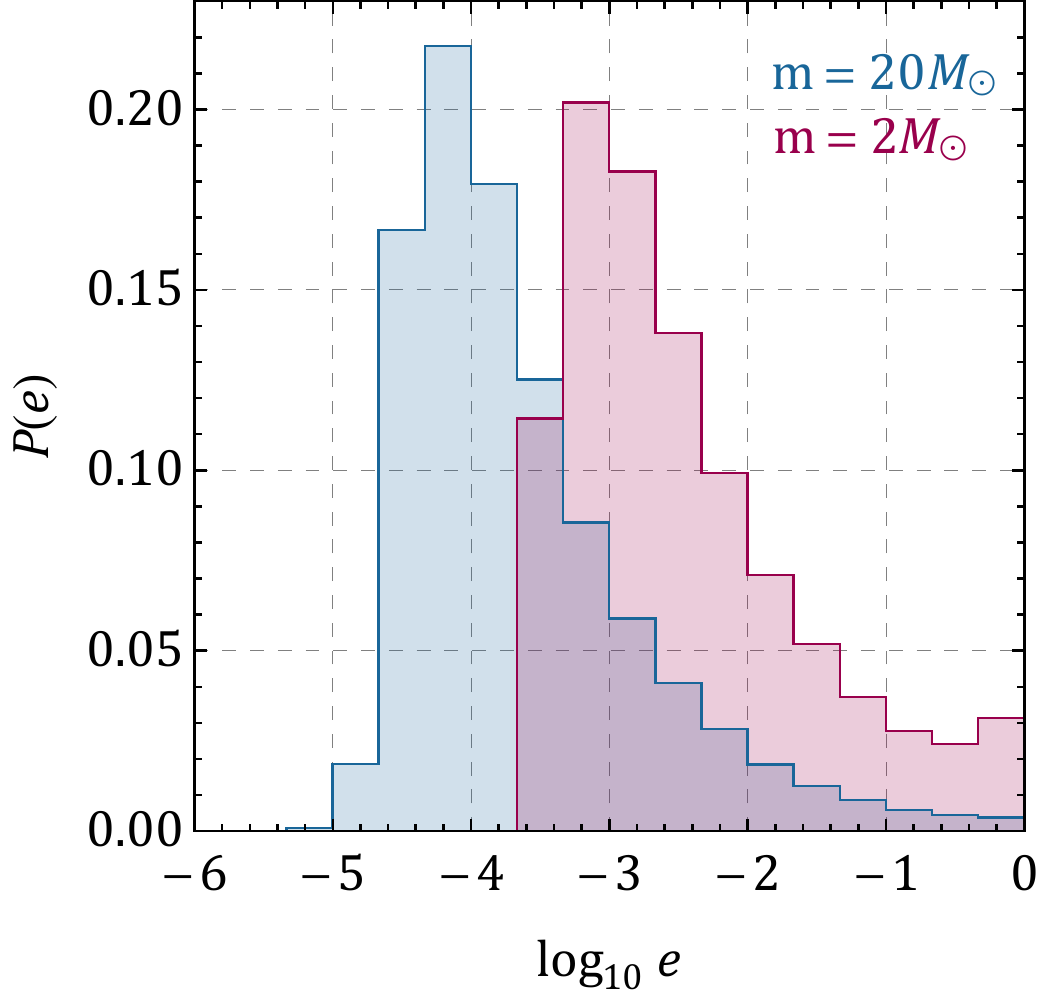}
\caption{The distribution of eccentricity $e_\text{LIGO}$ of binaries in a SMBH-carrying NC. For blue region, the initial conditions are chosen to be $m_0=m_1=10M_\odot$, $m_2=4\times 10^{6}M_\odot$. The stellar background profile is Bahcall-Wolf cusp $\al=7/4$ with mass segregation in BBH components. The red region in the left panel corresponds to replacing the cusp profile by a core profile with $\al=0.5$ and without mass segregation. The red region in the right panel corresponds to the cusp profile with $m=2M_\odot$.}
\label{fig_eHisto}
\end{figure}

\section{Discussion}
\label{sec_discussions}

We expect to observe inspiraling black hole binaries  frequently in future GW detections, which might put us in a position to study their formation when more statistics are available.

In this paper we further advanced the idea that the orbital eccentricity could be an important parameter for understanding the formation channels of inspiral binaries, and developed a more complete analytical understanding of the eccentricity distribution of binaries in galactic nuclei. This allows us to map parameters determined by BBH environments to eccentricity distributions directly, without using the tidal sphere of influence (and the associated function we had defined as $f(e)$) or numerical simulations. The statistical distribution of very small eccentricity $e\ll 1$ can contain very important information and might provide a unique probe into the origin of binary black hole mergers and to the surrounding density distributions.

In our analytical study of binary evolution, we have accounted for the perturbation from the third body to quadrupole order, the post-Newtonian precession of the orbit, and the back reaction of emitted gravitational waves. We see a nice agreement between the analytical estimate presented here and the numerical results presented in \cite{Antonini:2012ad}. The analytical framework we developed can be useful to study more efficiently the effect of various initial conditions that are relevant to the formation of binaries. As an example, we show that there is an anti-correlation between the binary mass and eccentricity for binary mergers in NCs, in contrast to expectations from binaries in GCs \cite{Wen:2002km}. 

Our analytical estimate works very well for small eccentricity, while it becomes less accurate for large eccentricity $e>0.1$,  where the three effects --- KL oscillation, PN precession, GW back reaction --- are equally important during the whole history of the binary evolution. This region can however be studied numerically as it is only a limited portion of the existing volume. In addition, we note that while the large-eccentricity events generally merge very quickly, their event rate is determined by the rate of replenishment, which turns out to be slow \cite{Antonini:2012ad} so we expect only a small fraction of events from the inner regions where large eccentricity might be generated. It will be interesting in the future to try to extend our approach to the octupole perturbation of the third body and also to other non-secular corrections to doubly averaged Hamiltonian, which could further boost the eccentricity generation. Finally, our method can also apply to other formation channels involving hierarchical triples such as in GCs and in the field. We leave these questions to future studies.

\paragraph{Acknowledgement.} In Memoriam of Yoshihide Kozai (1928 -- 2018). We thank Imre Bartos, Evgeny Grishin, Savvas Koushiappas, Erez Michaely, Hagai Perets, Sterl Phinney, Johan Samsing for helpful conversations. LR is supported by an NSF grant PHY-1620806, a Kavli Foundation grant ``Kavli Dream Team'', and a Simons Foundation grant 511879. ZZX is supported in part by Center of Mathematical Sciences and Applications, Harvard University.

\begin{appendix}

\section{LIGO Detectability of Small Eccentricities}
\label{app_small_e}

GWs produced by elliptical binaries present several new features compared with the ones from circular binaries, including the modified waveform, the appearance of higher harmonics, and also the shifted peak value of frequency. For very eccentric orbits $e\sim 1$ we may be able to see a wide range of harmonics peaked at a much higher frequency than the orbital frequency, and for mildly eccentric orbits $0.1\lesssim e\lesssim 0.9$, we expect to see one or several of higher harmonics, with amplitude smaller than or comparable with the base frequency. On the other hand, if the eccentricity is small, $e<0.1$, the amplitude of the higher harmonics would be too small to be visible, and in this case we hope to detect the ellipticity by monitoring the modified waveform. In this appendix, we estimate the sensitivity of LIGO to small eccentricities. See \cite{Huerta:2016rwp,Huerta:2017kez} for waveform models of eccentric binary mergers in LIGO.

The waveform is governed by the orbital frequency as a function of time, $\omega=\omega(t)$, and thus by the semi-major axis $a=a(t)$ since the two are related by $\omega^2=Gm/a^3$. It is known that $a=a(t)$ is governed by,
\bge
\label{dadtGW}
  \FR{\di a}{\di t}=-\FR{64G^3\mu m^2}{5c^5a^3}\FR{1}{(1-e^2)^{7/2}}\bigg(1+\FR{73}{24}e^2+\FR{37}{96}e^4\bigg)\simeq -\FR{64G^3\mu m^2}{5c^5a^3}\bigg(1+\FR{157}{24}e^2\bigg),
\ede
where we have expanded the formula around $e=0$. It is also known that $a$ and $e$ are related to each other by $a/a_\text{LIGO}=g(e)/g(e_\text{LIGO})$ with $a_\text{LIGO}$ and $e_\text{LIGO}$ evaluated at the time of entering the LIGO window, and,
\bge
\label{ge}
  g(e)=\FR{e^{12/19}}{1-e^2}\bigg(1+\FR{121}{304}e^2\bigg)^{870/2299}\simeq e^{12/19}.
\ede
Therefore, we can find $f_\text{GW}(t)=\omega(t)/\pi$ for small $e_0$ by integrating (\ref{dadtGW}) together with (\ref{ge}),
\begin{align}
  f_\text{GW}(t)
  \simeq &~151\text{Hz}\bigg(\FR{m_c}{M_\odot}\bigg)^{-5/8}\bigg(\FR{t}{1\text{s}}\bigg)^{-3/8} -2.11\text{Hz}\bigg(\FR{m_c}{M_\odot}\bigg)^{25/36}\bigg(\FR{t}{1\text{s}}\bigg)^{5/12}e_{\text{LIGO}}^2,
\end{align}
where $m_c=m^{2/5}\mu^{3/5}$ is the so-called chirp mass. In this expression we have taken the lower limit of the LIGO frequency window to be $10$Hz, and thus $a_\text{LIGO}\simeq 513\text{km}\times(m/M_\odot)^{1/3}$ for small eccentricities.

When $e_\text{LIGO}$ is very small, the correction from the second term of the above equation can be recognized by tracing the change in the number of periods $N$ of GWs from the time of entering the LIGO band to the time of coalescence. Requiring that $|\Delta N|\geq 1$ as a rough criterion of LIGO detectability, we find,
\bge
  e_\text{LIGO}\gtrsim 0.0049\bigg(\FR{m_c}{M_\odot}\bigg)^{5/6}.
\ede

\end{appendix}

%\end{fmffile}
\end{document}